\documentclass[11pt,draftcls,onecolumn]{IEEEtran}

\usepackage{amsfonts}
\usepackage{amsbsy}
\usepackage{amssymb}
\usepackage{amscd}
\usepackage[cmex10]{amsmath}
\usepackage{graphicx}
\usepackage[usenames,dvipsnames]{pstricks}

\newtheorem{definition}{Definition}
\newtheorem{assumption}{Assumption}

\newtheorem{theorem}{Theorem}
\newtheorem{lemma}{Lemma}

\newtheorem{corollary}{Corollary}

\newcommand{\mat}[1]{\boldsymbol{#1}}
\newcommand{\tr}[1]{\text{tr}\left( #1\right)}
\newcommand{\rank}[1]{\text{rank}\left( #1\right)}

\newcommand{\pp}[1]{{\left( #1 \right)}}
\newcommand{\br}[1]{{\left\{ #1 \right\}}}
\newcommand{\norm}[1]{{ \left\Vert #1 \right\Vert }}
\newcommand{\abs}[1]{{ \left| #1 \right| }}
\newcommand{\sabs}[1]{{ \left| #1 \right|^2 }}
\newcommand{\snorm}[1]{{ \left\Vert #1 \right\Vert^2 }}
\newcommand{\dom}[1]{\geq^{#1}}

\def\hf{{\frac{1}{2}}}

\def\H{{H}} 
\def\bI{{\mat{I}}} 
\def\bQ{{\mat{Q}}} 
\def\bU{{\mat{U}}} 
\def\bh{{\mat{h}}} 
\def\bx{{\mat{x}}} 
\def\bz{{\mat{z}}} 
\def\bw{{\mat{w}}} 
\def\bv{{\mat{v}}} 

\begin{document}

%
\title{Optimal Beamforming in Interference Networks with Perfect Local Channel Information\thanks{Part of this work has been performed in the framework of the European research project SAPHYRE, which is partly funded by the European Union under its FP7 ICT Objective 1.1 - The Network of the Future. This work is also supported in part by the Deutsche Forschungsgemeinschaft (DFG) under grant Jo 801/4-1.}}%
%
%
\author{Rami~Mochaourab,~\IEEEmembership{Student~Member,~IEEE,}
        and~Eduard~Jorswieck,~\IEEEmembership{Senior~Member,~IEEE}%
\thanks{The authors are with the Department of Electrical Engineering and Information Technology, Dresden University of Technology, 01062 Dresden, Germany. E-mail: \{Rami.Mochaourab,Eduard.Jorswieck\}@tu-dresden.de. Phone: +49-351-46332239. Fax: +49-351-46337236.}
\thanks{Part of this work has been presented at IEEE International Zurich Seminar on Communications, Zurich, Switzerland, March 3--5, 2010 \cite{Jorswieck2010a} and International Workshop on Cognitive Information Processing, Elba Island, Italy, June 14--17, 2010 \cite{Jorswieck2010b}.}}
%
\maketitle
\begin{abstract}
We consider settings in which $T$ multi-antenna transmitters and $K$ single-antenna receivers concurrently utilize the available communication resources. Each transmitter sends useful information only to its intended receivers and can degrade the performance of unintended systems. Here, we assume the performance measures associated with each receiver are monotonic with the received power gains. In general, the systems' joint operation is desired to be Pareto optimal. However, designing Pareto optimal resource allocation schemes is known to be difficult. In order to reduce the complexity of achieving efficient operating points, we show that it is sufficient to consider rank-1 transmit covariance matrices and propose a framework for determining the efficient beamforming vectors. These beamforming vectors are thereby also parameterized by $T(K-1)$ real-valued parameters each between zero and one. The framework is based on analyzing each transmitter's power gain-region which is composed of all jointly achievable power gains at the receivers. The efficient beamforming vectors are on a specific boundary section of the power gain-region, and in certain scenarios it is shown that it is necessary to perform additional power allocation on the beamforming vectors. Two examples which include broadcast and multicast data as well as a cognitive radio application scenario illustrate the results.
\end{abstract}
%
%
%
\section{Introduction}

%
Interference is known to be one of the major factors that limits the performance of a communication system in a wireless network. This situation is common in multiuser settings when the systems concurrently share the available communication resources. In general interference networks, the performance measure of individual users is described by a utility function. This function depends in a monotonic way on the received signal power, interference signal power and noise power. The joint operation of the systems is efficient if it is not possible to improve the performance of one system without degrading the performance of another. In this case, the operating point is said to be Pareto optimal. It is always desired to design resource allocation schemes that lead to Pareto optimal operation points. In this way, the available communication resources are utilized efficiently to grant efficient operation of the systems. However, developing efficient resource allocation schemes is not straightforward and proves to be difficult. For instance, the problem of finding the maximum sum-rate or the proportional-fair operating point in the multiple-input single-output (MISO) interference channel (IC) is proven to be strongly NP-hard\footnote{Interestingly, these problems are efficiently solvable if rate requirements or interference constraints on each system are fixed.} \cite{Liu2010}.

In a multiuser setting, efficient operation of the systems requires the transmitters to maximize the power gain at intended receivers and also minimize the power gain at unintended receivers. In this work, we characterize the transmission strategies of each transmitter that are relevant to achieve Pareto optimal operating points. Moreover, we parameterize these by real values between zero and one. In this way, the set of efficient transmission strategies is confined and represented by low dimensional real parameters. This result tremendously reduces the complexity of designing efficient resource allocation schemes, and the parametrization can be utilized for low complexity coordination between transmitters.

%
%
We give a brief reference to related work in the MISO IC and a few of their applications. The MISO IC is an example of an interference network where the systems consist of transmitter-receiver pairs. For the two-user case, real-valued parametrization of each transmitter's efficient beamforming vectors is provided in \cite{Jorswieck2008b}. The beamforming vectors that achieve Pareto optimal points are proven to be a linear combination of zero-forcing (ZF) transmission and maximum ratio transmission (MRT). Based on this characterization, a monotonic optimization framework is developed in \cite{Jorswieck2009} to find maximum sum-rate, proportional-fair and minimax operating points. The parametrization in \cite{Jorswieck2008b} relates to a parametrization using the virtual SINR framework in \cite{Zakhour2009}. The use of this framework is motivated by the design of distributed algorithms that require local channel state information (CSI) at each transmitter. This framework is extended to the precoding design in MIMO settings in \cite{Zakhour2009a}. The concept of combining the MRT and ZF strategies is important for developing so-called distributed bargaining algorithms. These algorithms improve the operation of the systems from the noncooperative outcome \cite{Ho2008, Lindblom2010}. In \cite{Ho2008}, a distributed bargaining algorithm is developed which requires one bit signaling between the transmitters. Extension to the precoding design in the multiple-input multiple-output (MIMO) case is given in \cite{Ho2010}. In \cite{Lindblom2010}, a similar distributed beamforming algorithm in the MISO IC is proposed for the case of statistical CSI at the transmitters. Also utilizing the parametrization in \cite{Jorswieck2008b}, a distributed bargaining process is proposed in \cite{Mochaourab2010} which requires four bit signaling between the transmitters. The process is proven to converge to an operating point arbitrarily close to the Pareto boundary and dominates the noncooperative outcome of the systems. In \cite{Lebrun2009}, the high signal to interference plus noise ratio (SINR) approximation of the achievable sum-rate of a system pair is utilized to determine suboptimal joint transmission strategies. The achieved performance is shown to be better than the joint MRT and joint ZF strategies.

In the $K$-user MISO IC, complex-valued parametrization of the Pareto boundary of the MISO IC rate-region is derived in \cite{Jorswieck2008} which requires $K(K-1)$ complex-valued parameters in order to attain all Pareto optimal points. In \cite{Shang2009}, the $K$-user MISO IC is considered with the capabilities of time sharing the resources between the links. All points on the Pareto boundary of the MISO IC rate region are achieved with $K(K-1)$ real valued parameters each between $0$ and $\pi$. In \cite{Zhang2010a}, the authors characterize the Pareto boundary of the MISO IC through controlling interference temperature constraints (ITC) at the receivers. Each Pareto optimal rate tuple is achieved iteratively when each transmitter optimizes its transmission constrained by the ITCs. It is shown that $K(K-1)$ real valued parameters, each between zero and a value depending on the channel vectors, are needed to achieve all Pareto optimal points. ITC is a terminology used in cognitive radio scenarios under the underlay paradigm \cite{Goldsmith2009}. It quantifies the amount of interference from the secondary transmitters that is tolerated by the primary users.

In \cite{Bjornson2010}, joint linear precoding is investigated taking into account the signaling overhead between the transmitters. The rate-region achieved with joint precoding is larger than the MISO IC rate-region, and all Pareto optimal beamforming vectors are parameterized by $K(K-1)$ complex-valued parameters. For the same setting, a recent result in \cite{Bjoernson} reduces the number of parameters to $K+L$ real-valued scalars, each between zero and one, where $L$ is the number of linear constraints on the transmission. Linear precoding MIMO IC algorithms are moreover investigated in \cite{Chae2010} for a two-user system.

%
%
While the above mentioned results are provided for the MISO IC setting, the parametrization of efficient transmission strategies in a general multiuser setting is not straightforward. Moreover, neither the ITC-based \cite{Zhang2010a} nor the Lagrangian-based \cite{Shang2009,Bjoernson} characterizations can be generalized to our framework. A further example of a MISO multiuser setting which can be applied to our framework is when a single transmitter sends common information to $K$ single-antenna receivers. This setting corresponds to multicast transmission. Since the transmission rate depends on the weakest link in the system, the transmitter optimizes its transmission to achieve max-min-fairness at the receivers \cite{Gershman2010}. The multicast beamforming problem to achieve max-min-fairness is proven to be NP hard for $K \geq N$ \cite{Sidiropoulos2006}, where $N$ is the number of transmit antennas. In \cite{Tomecki2010}, the two-user multicast max-min-fair problem is studied, and the set of beamforming vectors which includes the solution of the max-min-fair problem is characterized.

%
%
In this work, we provide a general framework for parameterizing the transmission strategies of each transmitter which are relevant to achieve Pareto optimal points. This framework is applicable to settings where the utility functions of the systems are monotonic in the received power gains. The contributions and outline of this paper are as follows:
\begin{itemize}
\item We investigate the properties of efficient transmission of a single transmitter. These properties are acquired on studying the transmitter's power gain-region (Section \ref{sec:powergain}). The power gain-region is composed of all jointly achievable power gains at the receivers. Of interest are the transmission strategies which achieve its boundary part in a specific direction. We prove that the boundary of the power gain-region is convex and always achieved with single-stream beamforming (Lemma \ref{thm:lemma_rank1}). Due to these properties, the corresponding strategies are characterized by real-valued parameters (Theorem \ref{theo:1}). Furthermore, we characterize under which conditions power control is needed for efficient transmission. (i) When the number of transmit antennas is greater than or equal to the number of receivers $K$ (Section \ref{sec:NgeqK}), we prove that full power transmission achieves all boundary points (Lemma \ref{thm:lemma_trace1}). In this case, $K-1$ real-valued parameters, each between zero and one, are needed to parameterize the beamforming vectors. (ii) When the number of transmit antennas is strictly less than the number of receivers (Section \ref{sec:NlessK}), we characterize the transmission strategies for which power control is needed. For this case, an additional real-valued parameter between zero and one is needed that varies the power level at the transmitter.

\item We utilize the developed single-transmitter framework for the multiple-transmitter case (Section \ref{sec:pareto}). Based on the network setting and the monotonicity properties of each receiver's utility function, the boundary part which is relevant for Pareto optimal operation is determined for each transmitter's gain-region. Consequently, each transmitter's efficient strategies are parameterized (Theorem \ref{thm:thm2}). We provide an example setting which includes broadcast and multicast data, and we apply the developed framework to this setting (Section \ref{sec:example}). Moreover, we apply the framework to the $K$-user MISO IC (Section \ref{sec:applications1}). As a special case, the result for the two-user MISO IC in \cite{Jorswieck2008} follows. In addition, we give an alternative characterization of the efficient transmission strategies (Corollary \ref{thm:cor1}) which is motivated by the application of null-shaping constraints in underlay cognitive radio scenarios (Section \ref{sec:applications2}). We prove that all Pareto optimal operating points can be characterized through the design of null-shaping constraints on noncooperative secondary transmitters. Extensions to the case of multiple antennas at the receivers is covered in Section \ref{sec:MIMO}.
\end{itemize}
\subsubsection*{Notations}%
Column vectors and matrices are given in lowercase and uppercase boldface letters, respectively. The notation $x_{k,\ell}$ describes the $\ell$th component of vector $\mat{x}_k$. The Euclidean norm of a vector $\mat{a}, \mat{a} \in \mathbb{C}^{N},$ is written as $\norm{\mat{a}}$, and the absolute value of  $b, b \in \mathbb{C},$ is $\abs{b}$. $(\cdot)^\H$ denotes the Hermitian transpose. The $i$th eigenvalue of a matrix $\mat{Z}$ is denoted by $\mu_i(\mat{Z})$. The eigenvector which belongs to the $i$th eigenvalue of the matrix $\mat{Z}$ is denoted by $\bv_{i}(\mat{Z})$. We always assume that the eigenvalues are ordered in nondecreasing order such that $\mu_i(\mat{Z}) \leq \mu_{i+1}(\mat{Z})$. Moreover, the eigenvectors corresponding to the largest and smallest eigenvalues of a matrix $\mat{Z}$ are specified as $\bv_{\text{max}}(\mat{Z})$ and $\bv_{\text{min}}(\mat{Z})$, respectively. The notation $\mat{Z} \succeq 0$ means that $\mat{Z}$ is positive semidefinite. The rank and trace of a matrix $\mat{Z}$ are given by $\rank{\mat{Z}}$ and $\tr{\mat{Z}}$, respectively. The orthogonal projector onto the column space of $\mat{Z}$ is $\mat{\Pi}_{Z} := \mat{Z}\pp{\mat{Z}^\H\mat{Z}}^{-1}\mat{Z}^\H$. The orthogonal projector onto the orthogonal complement of the column space of $\mat{Z}$ is $\mat{\Pi}_{Z}^{\perp} := \bI - \mat{\Pi}_{Z}$, where $\bI$ is an identity matrix. $\mathbb{E}\pp{\cdot}$ denotes statistical expectation. The set of non-negative real numbers is $\mathbb{R}_+$. The cardinality of a set $\mathcal{K}$ is written as $\abs{\mathcal{K}}$.

\section{System and Channel Model}\label{sec:SysChan}%
We consider $T$ transmitters and $K$ receivers sharing the same spectral band. Define the set of transmitters as $\mathcal{T} := \{1,...,T\}$ and receivers as $\mathcal{K} := \{1,...,K\}$. Each transmitter sends useful information to at least one receiver. For transmitter $k, k \in \mathcal{T}$, let $\overline{\mathcal{K}}(k)\subseteq \mathcal{K}$ denote the set of its intended receivers for which useful information is sent to, and let $\underline{\mathcal{K}}(k)=\mathcal{K}\backslash \overline{\mathcal{K}}(k)$ be the set of its unintended receivers. Each transmitter $k$ is equipped with $N_k$ antennas, and each receiver with a single antenna. The quasi-static block flat-fading instantaneous channel vector from transmitter $k, k \in \mathcal{T},$ to receiver $\ell, \ell \in \mathcal{K},$ is denoted by $\mat{h}_{k\ell} \in \mathbb{C}^{N_k \times 1}$. The transmit covariance matrix of transmitter $k$ is given as $\mat{Q}_{k} \in \mathbb{C}^{N_k \times N_k}$, $\bQ_k \succeq 0$. We do not make any assumptions on the number of data streams applied at the transmitters. The basic model for the matched-filtered, symbol-sampled complex baseband data received at receiver $\ell$ is
\begin{eqnarray}
  y_\ell=\sum_{k=1}^T \mat{h}_{k\ell}^\H \mat{Q}^{\hf}_k \mat{s}_k + n_\ell, \label{eq:systemmodel}
\end{eqnarray}
\noindent where $\mat{s}_k$ is the symbols vector transmitted by transmitter $k$ and $n_\ell$ are the noise terms which we model as independent and identically distributed (i.i.d.)~complex Gaussian with zero mean and variance $\sigma^2$. Each transmitter has a total power constraint of $P := 1$ which leads to the constraint $\tr{\mat{Q}_{k}} \leq 1$, $k \in \mathcal{T}$. Throughout, we define the signal to noise ratio (SNR) as $1/\sigma^2$. The feasible set of covariance matrices for transmitter $k$ is defined as%
\begin{equation}\label{eq:S_set}
\mathcal{S}_k := \br{\bQ_k \in \mathbb{C}^{N_k \times N_k} : \bQ_k \succeq 0, \tr{\bQ_k} \leq 1}.
\end{equation}%
\noindent Note that $\mathcal{S}_k$ is compact and convex. We assume each transmitter has local CSI, i.e., it has perfect knowledge of the channel vectors only between itself and all receivers \cite{Zakhour2009a}. This ideal scenario serves as an upper bound to the more realistic case in which imperfect or partial CSI at the transmitters is available. Extensions in this direction are reported in \cite{Lindblom2009,Lindblom2010a}. In these works, Pareto efficient transmission strategies are characterized for the two-user MISO IC with partial CSI at the transmitters.

\subsection{Assumptions on Performance Measure}\label{sec:utility}
The performance measure of a system in an interference network is usually described by a utility function. The utility function associated with a receiver depends on the power gains originating from the transmitters in the network. Define the power gain achieved by transmitter $k$ at a receiver $\ell$ as
\begin{equation}\label{eq:pgain}
x_{k,\ell}(\bQ_k) = \mat{h}_{k\ell}^H \mat{Q_k} \mat{h}_{k\ell},
\end{equation}
where $x_{\ell}(\bQ_k) \in \mathbb{R}_+$ since $\bQ_k$ is positive semidefinite. The utility function associated with a receiver $\ell$ is defined as $u_\ell : \mathbb{R}_+^{T} \rightarrow \mathbb{R}_+$, where $T$ is the number of transmitters in the network.

\begin{assumption}\label{def:utility} The utility function $u_\ell, \ell \in \mathcal{K},$ has the following properties:
\begin{itemize}
\item[A.] If $\ell \in \mathcal{\overline{K}}(k),$ then $u_\ell$ is monotonically increasing in the power gain from transmitter $k$, i.e.,
\begin{equation}\label{def:utility_A}
u_\ell\pp{x_{1,\ell}\pp{\bQ_1},...,x_{T,\ell}\pp{\bQ_T}} \leq u_\ell\pp{x_{1,\ell}\pp{\bQ_1},..., x_{k,\ell}(\widehat{\bQ}_k),...,x_{T,\ell}\pp{\bQ_T}},
\end{equation}
for $x_{k,\ell}\pp{\bQ_1} \leq x_{k,\ell}(\widehat{\bQ}_k)$.
\item[B.] If $\ell \in \mathcal{\underline{K}}(k),$ then $u_\ell$ is monotonically decreasing in the power gain from transmitter $k$, i.e.,
\begin{equation}\label{def:utility_B}
u_\ell\pp{x_{1,\ell}\pp{\bQ_1},...,x_{T,\ell}\pp{\bQ_T}} \geq u_\ell\pp{x_{1,\ell}\pp{\bQ_1},..., x_{k,\ell}(\widehat{\bQ}_k),...,x_{T,\ell}\pp{\bQ_T}},
\end{equation}
for $x_{k,\ell}\pp{\bQ_k} \leq x_{k,\ell}(\widehat{\bQ}_k)$.$\hfill\Box$
\end{itemize}
\end{assumption}

Assumption \ref{def:utility} describes the settings where the performance measure at a receiver increases monotonically with increased power gain from intended transmitters and decreases monotonically with increased power gain from unintended transmitters. An example utility function which satisfies Assumption \ref{def:utility} is the signal to interference plus noise ratio (SINR).

The \emph{utility region} is the set of all achievable utility tuples defined as:
\begin{eqnarray}\label{eq:utilityregion}
  \mathcal{U} := \br{ \pp{u_1(x_{1,1}\pp{\bQ_1},\ldots,x_{T,1}\pp{\bQ_T}),\ldots, u_K(x_{1,K}\pp{\bQ_1},\ldots,x_{T,K}\pp{\bQ_T})} : \bQ_{k} \in \mathcal{S}_k, k \in \mathcal{T} } \subset \mathbb{R}_+^K \label{eq:all}.
\end{eqnarray}
\noindent The efficient operating points in the utility region correspond to those in which it is impossible to improve the performance of one system without simultaneously degrading the performance of at least one other system. Such operating points are called Pareto optimal and are defined formally as follows.
\begin{definition}
A tuple $(u_1,...,u_K) \in \mathcal{U}$ is Pareto optimal if there is no other tuple $({u'}_1,...,{u'}_K) \in \mathcal{U}$ such that $({u'}_1,...,{u'}_K) \geq (u_1,...,u_K)$, where the inequality is component-wise and strict for at least one component. The set of all Pareto optimal operating points constitutes the \emph{Pareto boundary} ($\mathcal{PB}$) of $\mathcal{U}$. $\hfill\Box$
\label{def:1}
\end{definition}
Next, we give an example setting where the systems' utility functions satisfy Assumption \ref{def:utility}.
\subsection{Example Setting}\label{sec:example_setting}
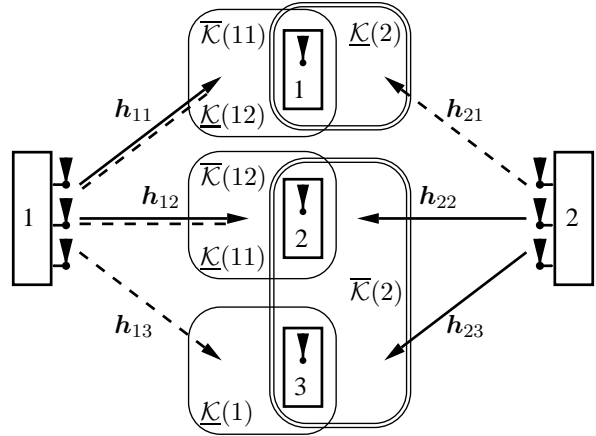
\begin{figure}[t]
\centering
\scalebox{0.9} 
{
\begin{pspicture}(0,-3.2)(8.6,3.2)
\psframe[linewidth=0.02,framearc=0.5,dimen=outer,doubleline=true,doublesep=0.04,doublecolor=white](5.9,0.9)(3.8,-3.1)
\psframe[linewidth=0.02,framearc=0.5,dimen=outer](4.8,-1.3)(2.6,-3.2)
\psframe[linewidth=0.02,framearc=0.5,dimen=outer,doubleline=true,doublesep=0.04,doublecolor=white](5.9,3.2)(3.8,1.3)
\psframe[linewidth=0.04,dimen=outer](0.6,1.0)(0.0,-1.0)
\psframe[linewidth=0.04,dimen=outer](8.6,1.0)(8.0,-1.0)
\psframe[linewidth=0.04,dimen=outer](4.6,2.8)(4.0,1.6)
\psframe[linewidth=0.04,dimen=outer](4.6,0.6)(4.0,-0.6)
\psframe[linewidth=0.04,dimen=outer](4.6,-1.6)(4.0,-2.8)
\psline[linewidth=0.04cm,linestyle=dashed,dash=0.16cm 0.16cm,arrowsize=0.04cm 3.0,arrowlength=2.0,arrowinset=0.0]{->}(7.6,0.5)(5.5,2.1)
\psline[linewidth=0.04cm,arrowsize=0.04cm 3.0,arrowlength=2.0,arrowinset=0.0]{->}(7.6,0.0)(5.1,0.0)
\psline[linewidth=0.04cm,arrowsize=0.04cm 3.0,arrowlength=2.0,arrowinset=0.0]{->}(7.6,-0.5)(5.5,-2.1)
\psline[linewidth=0.04cm,arrowsize=0.04cm 3.0,arrowlength=2.0,arrowinset=0.0]{->}(1.0,0.5)(3.1,2.1)
\psline[linewidth=0.04cm,arrowsize=0.04cm 3.0,arrowlength=2.0,arrowinset=0.0]{->}(1.0,0.0)(3.5,0.0)
\psline[linewidth=0.04cm,linestyle=dashed,dash=0.16cm 0.16cm,arrowsize=0.04cm 3.0,arrowlength=2.0,arrowinset=0.0]{->}(1.0,-0.5)(3.1,-2.1)
\usefont{T1}{ptm}{m}{n}
\rput(0.246875,0.01){1}
\usefont{T1}{ptm}{m}{n}
\rput(4.246875,1.91){1}
\usefont{T1}{ptm}{m}{n}
\rput(8.278594,0.01){2}
\usefont{T1}{ptm}{m}{n}
\rput(4.2785935,-0.29){2}
\usefont{T1}{ptm}{m}{n}
\rput(4.2676563,-2.49){3}
\usefont{T1}{ptm}{m}{n}
\rput(1.8014063,1.61){$\mat{h}_{11}$}
\usefont{T1}{ptm}{m}{n}
\rput(2.2014062,0.31){$\mat{h}_{12}$}
\usefont{T1}{ptm}{m}{n}
\rput(1.8014063,-1.59){$\mat{h}_{13}$}
\usefont{T1}{ptm}{m}{n}
\rput(6.7014065,1.61){$\mat{h}_{21}$}
\usefont{T1}{ptm}{m}{n}
\rput(6.3014064,0.31){$\mat{h}_{22}$}
\usefont{T1}{ptm}{m}{n}
\rput(6.7014065,-1.59){$\mat{h}_{23}$}
\psline[linewidth=0.04cm,fillcolor=black,dotsize=0.04cm 1.5,arrowsize=0.06cm 3.0,arrowlength=2.0,arrowinset=0.0]{*-<}(4.3,2.3)(4.3,2.7)
\psline[linewidth=0.04cm,fillcolor=black,dotsize=0.04cm 2.0,arrowsize=0.08cm 3.0,arrowlength=2.0,arrowinset=0.0]{*-<}(0.8,0.5)(0.8,0.9)
\psline[linewidth=0.04cm,fillcolor=black,dotsize=0.04cm 2.0,arrowsize=0.08cm 3.0,arrowlength=2.0,arrowinset=0.0]{*-<}(0.8,-0.1)(0.8,0.3)
\psline[linewidth=0.04cm,fillcolor=black,dotsize=0.04cm 1.5,arrowsize=0.06cm 3.0,arrowlength=2.0,arrowinset=0.0]{*-<}(4.3,0.1)(4.3,0.5)
\psline[linewidth=0.04cm,fillcolor=black,dotsize=0.04cm 1.5,arrowsize=0.06cm 3.0,arrowlength=2.0,arrowinset=0.0]{*-<}(4.3,-2.1)(4.3,-1.7)
\psline[linewidth=0.04cm,fillcolor=black,dotsize=0.04cm 2.0,arrowsize=0.08cm 3.0,arrowlength=2.0,arrowinset=0.0]{*-<}(7.8,0.5)(7.8,0.9)
\psline[linewidth=0.04cm,fillcolor=black,dotsize=0.04cm 2.0,arrowsize=0.08cm 3.0,arrowlength=2.0,arrowinset=0.0]{*-<}(7.8,-0.1)(7.8,0.3)
\psline[linewidth=0.04cm]{cc-cc}(7.8,-0.1)(7.9933352,-0.1)
\usefont{T1}{ptm}{m}{n}
\rput(3.1814063,-2.89){$\underline{\mathcal{K}}(1)$}
\usefont{T1}{ptm}{m}{n}
\rput(5.3814063,2.71){$\underline{\mathcal{K}}(2)$}
\usefont{T1}{ptm}{m}{n}
\rput(3.2814062,0.61){$\overline{\mathcal{K}}(12)$}
\usefont{T1}{ptm}{m}{n}
\rput(5.391406,-1.09){$\overline{\mathcal{K}}(2)$}
\psline[linewidth=0.04cm]{cc-cc}(0.8,0.5)(0.6,0.5)
\psline[linewidth=0.04cm]{cc-cc}(0.8,-0.1)(0.6,-0.1)
\psline[linewidth=0.04cm]{cc-cc}(7.8,0.5)(8.0,0.5)
\psline[linewidth=0.04cm,fillcolor=black,dotsize=0.04cm 2.0,arrowsize=0.08cm 3.0,arrowlength=2.0,arrowinset=0.0]{*-<}(0.8,-0.7)(0.8,-0.3)
\psline[linewidth=0.04cm]{cc-cc}(0.8,-0.7)(0.6,-0.7)
\psline[linewidth=0.04cm,fillcolor=black,dotsize=0.04cm 2.0,arrowsize=0.08cm 3.0,arrowlength=2.0,arrowinset=0.0]{*-<}(7.8,-0.7)(7.8,-0.3)
\psline[linewidth=0.04cm]{cc-cc}(7.8,-0.7)(7.9933352,-0.7)
\psframe[linewidth=0.02,framearc=0.5,dimen=outer](4.8,3.1)(2.6,1.2)
\psframe[linewidth=0.02,framearc=0.5,dimen=outer](4.8,1.0)(2.6,-0.9)
\usefont{T1}{ptm}{m}{n}
\rput(3.2714062,-0.59){$\underline{\mathcal{K}}(11)$}
\usefont{T1}{ptm}{m}{n}
\rput(3.2814062,2.71){$\overline{\mathcal{K}}(11)$}
\usefont{T1}{ptm}{m}{n}
\rput(3.2714062,1.51){$\underline{\mathcal{K}}(12)$}
\psline[linewidth=0.04cm,linestyle=dashed,dash=0.16cm 0.16cm](1.04,0.44)(2.88,1.84)
\psline[linewidth=0.04cm,linestyle=dashed,dash=0.16cm 0.16cm](1.0,-0.08)(3.16,-0.08)
\end{pspicture}
}
\caption{An example setting for the described system model. There exist two transmitters, each equipped with three antennas, and three single antenna receivers. The solid arrows refer to the intended receivers of a transmitter, while the dashed arrows refer to interference directions.}
\label{fig:Kset}
\end{figure}

%
Consider two transmitters each using three transmit antennas, and three single antenna receivers as depicted in \figurename~\ref{fig:Kset}. The operation of the systems is as follows:
\begin{itemize}
\item Broadcast Channel (BC): Transmitter $1$ transmits different useful data to receivers $1$ and $2$ simultaneously. We assume transmitter $1$ chooses the transmit covariance matrices $\bQ_{11}$ with $\tr{\bQ_{11}} = p_{11}$ for receiver $1$ and $\bQ_{12}$ with $\tr{\bQ_{12}} = p_{12}$ for receiver $2$. Hence, transmitter $1$ can be considered as two virtual transmitters\footnote{This transmission strategy is suboptimal, however less complex and more robust than dirty paper coding \cite{Weingarten2004}.}, $11$ and $12$, coupled by the total power constraint, $p_{11} + p_{12} \leq 1$. The receivers are identified in the following receiver sets: ${1} \in \overline{\mathcal{K}}(11), {1} \in \underline{\mathcal{K}}(12)$, ${2} \in \overline{\mathcal{K}}(12), {2} \in \underline{\mathcal{K}}(11)$.

\item Multiple Access Channel (MAC): Transmitters $12$ and $2$ send distinct useful information to receiver $2$. Receiver $2$ decodes the data from transmitter $12$ and $2$ successively. Thus, ${2} \in \overline{\mathcal{K}}(12), {2} \in \overline{\mathcal{K}}(2)$.

\item Multicast: Transmitter $2$ sends common useful data in a multicast to receivers $2$ and $3$. The receivers are identified in the following receiver sets: ${2} \in \overline{\mathcal{K}}(2), {3} \in \overline{\mathcal{K}}(2)$.

\item Interference Channel (IC): Transmitter $2$ induces interference on receiver $1$, while transmitter $1$ induces interference on receiver $3$.
\end{itemize}

The receiver sets are summarized in \figurename~\ref{fig:Kset}, and the solid and dashed arrows refer to useful and not useful signal directions, respectively. The achievable rate at receiver $1$ is
\begin{equation}\label{eq:utility1}
u_1 \pp{x_{11,1}\pp{\bQ_{11}},x_{12,1}\pp{\bQ_{12}},x_{2,1}\pp{\bQ_{2}}} = \log_2 \pp{1 + \frac{ \bh_{11}^\H \bQ_{11}\bh_{11} }{\sigma^2 + {\bh_{11}^\H \bQ_{12} \bh_{11}} + {\bh_{21}^\H \bQ_{2} \bh_{21}}}},
\end{equation}%
\noindent which is monotonically increasing in $x_{11,1}\pp{\bQ_{11}}$ and monotonically decreasing in the power gains from transmitters $12$ and $2$. The utility at receiver $2$ is its sum capacity \cite{Tse2005},
\begin{equation}\label{eq:utility2}
u_2 \pp{x_{11,2}\pp{\bQ_{11}},x_{12,2}\pp{\bQ_{12}},x_{2,2}\pp{\bQ_{2}}} = \log_2 \pp{1 + \frac{{\bh_{12}^\H \bQ_{12} \bh_{12}} + {\bh_{22}^\H \bQ_{2} \bh_{22}}}{\sigma^2 + {\bh_{12}^\H \bQ_{11} \bh_{12}}}},
\end{equation}%
\noindent which is monotonically increasing in $x_{12,2}\pp{\bQ_{12}}$ and $x_{2,2}\pp{\bQ_{2}}$. The utility function at receiver $3$ is the achievable rate\footnote{Note that the transmission rate at transmitter $2$ has to be chosen such that both receiver $2$ and $3$ can decode the data successfully. We do not consider this requirement in \eqref{eq:utility2} and \eqref{eq:utility3} since this is beyond the scope of this paper. However, these rates can be achieved using rateless coding \cite{Luby2002,MacKay2005}.},
\begin{equation}\label{eq:utility3}
u_3 \pp{x_{11,3}\pp{\bQ_{11}},x_{12,3}\pp{\bQ_{12}},x_{2,3}\pp{\bQ_{2}}} = \log_2 \pp{1 + \frac{ {\bh_{23}^\H \bQ_2 \bh_{23}}}{\sigma^2 + {\bh_{13}^\H \bQ_{11} \bh_{13}} + {\bh_{13}^\H \bQ_{12} \bh_{13}} }}.
\end{equation}
\noindent which is monotonically increasing in $x_{2,3}\pp{\bQ_{2}}$. The utility functions in \eqref{eq:utility1}-\eqref{eq:utility3} satisfy properties A and B in Assumption \ref{def:utility}. We return to this setting in Section \ref{sec:example} after we formalize the framework for efficient beamforming and resource allocation.

\section{Power Gain Region}\label{sec:powergain}%
In this section, a single transmitter $k, k \in \mathcal{T},$ is considered along with all $K$ receivers. The subscript $k$ in all terms referring to the single transmitter is omitted for convenience. For example, $\bh_\ell$ is written instead of $\bh_{k\ell}$. The receiver sets of the transmitter are written as $\overline{\mathcal{K}}$ and $\underline{\mathcal{K}}$ instead of $\overline{\mathcal{K}}(k)$ and $\underline{\mathcal{K}}(k)$. We return to include the indication to the transmitter in the next sections when multiple transmitters are considered. Here, we study the transmission effects of a single transmitter on all existing receivers. Thereby, we characterize its transmit covariance matrices that are relevant for its efficient operation in the multiuser system. For all feasible transmit covariance matrices, a power gain-region of a single transmitter is the set that consists of all joint power gains achievable at the receivers. The power gain-region of a transmitter is defined as
\begin{eqnarray}\label{eq:Omegak}
	\Omega := \left\{ \pp{x_{1}(\bQ),...,x_{K}(\bQ)} : \bQ \in \mathcal{S} \right\} \subset \mathbb{R}_+^K,
\end{eqnarray}
\noindent where $\mathcal{S}$ is defined in \eqref{eq:S_set}. Note that $x_{1}(\bQ),...,x_{K}(\bQ)$ are the main elements of the utility functions in \eqref{eq:utility1}-\eqref{eq:utility3}. An important property of the gain-region $\Omega$ is its convexity. Having this property is convenient for characterizing the points on its boundary using simple programming problems based on the Hyperplane Separation theorem \cite[Theorem 1.3]{Tuy1998}.
\begin{lemma}\label{thm:lemma_conv}
The set $\Omega$ in \eqref{eq:Omegak} is a compact and convex set.
\end{lemma}
\begin{IEEEproof}
The proof is provided in Appendix \ref{proof:lemma_conv}.
\end{IEEEproof}

Points that lie on the boundary of the gain-region $\Omega$ are of interest since these points characterize extreme power gains achievable at the receivers. At these points, the transmitter cannot increase the power gain in one direction of the gain-region without decreasing the power gain in any other direction. We give an example later in this section which further clarifies the importance of the boundary points of the gain-region for efficient operation of a transmitter. Next, we formalize the boundary of the set $\Omega$ following the definitions in \cite{Jorswieck2009}. There, these definitions were used to derive the solution of monotonic optimization problems \cite{Tuy2000,Tuy1998}.
%
%
\begin{definition}\label{def:uppBoundary}
	A point $\mat{y} \in \mathbb{R}_+^n$ is called upper boundary point of a compact convex set $\mathcal{C}$ if $\mat{y} \in \mathcal{C}$ while the set $\left\{ \mat{y}' \in \mathbb{R}_+^n : \mat{y}' > \mat{y} \right\} \subset \mathbb{R}_+^n \setminus \mathcal{C}$, where the inequality in $\mat{y}' > \mat{y}$ is componentwise. The set of upper boundary points of $\mathcal{C}$ is denoted by $\partial^+\mathcal{C}$.$\hfill\Box$
\end{definition}

Definition \ref{def:uppBoundary} describes only one boundary part of a compact convex set. The straightforward extension to describe all boundary parts of this set is to define its upper boundary in direction $\mat{e}, \mat{e} \in \{-1, +1\}^n$. For this purpose, we first need the following definition.
%
%
\begin{definition}\label{def:vect_dominates}
	A vector $\mat{x}$ dominates a vector $\mat{y}$ in direction $\mat{e}$, written as $\mat{x} \dom{\mat{e}} \mat{y}$, if $x_\ell e_\ell \geq y_\ell e_\ell$ for all $\ell$, $1\leq \ell \leq n,$ and the inequality has at least one strict inequality.$\hfill\Box$
\end{definition}
%
%
\begin{definition}
	A point $\mat{y} \in \mathbb{R}_+^n$ is called upper boundary point of a compact convex set $\mathcal{C}$ in direction $\mat{e}$ if $\mat{y} \in \mathcal{C}$ while the set $\left\{ \mat{y}' \in \mathbb{R}_+^n : \mat{y}' \dom{\mat{e}} \mat{y} \right\} \subset \mathbb{R}_+^n \setminus \mathcal{C}.$ We denote the set of upper boundary points in direction $\mat{e}$ as $\partial^{\mat{e}} \mathcal{C}$.$\hfill\Box$
\end{definition}

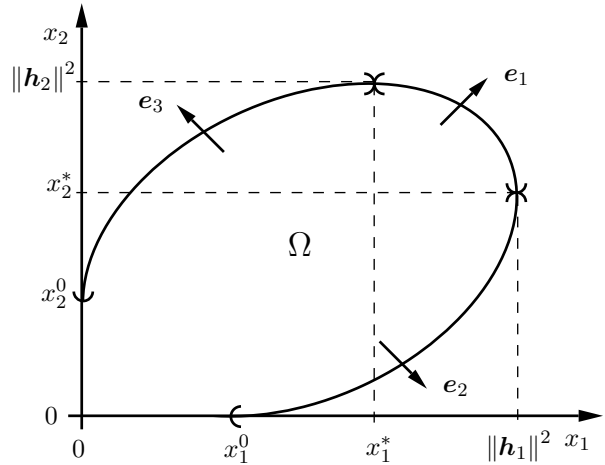
\begin{figure}[t]
\centering
\scalebox{0.9} 
{
\begin{pspicture}(0,-3.2655797)(9.602813,3.572233)
\rput{-335.073}(0.42731857,-2.1257062){\psellipse[linewidth=0.04,dimen=outer](5.0223765,-0.09618402)(3.4003057,2.2307498)}
\psframe[linewidth=0.0020,linecolor=white,dimen=outer,fillstyle=solid](4.002938,-0.84576696)(1.7989373,-2.5497673)
\psline[linewidth=0.04cm,arrowsize=0.06cm 3.0,arrowlength=2.0,arrowinset=0.0]{->}(1.8009375,-2.747767)(1.8009375,3.552233)
\usefont{T1}{ptm}{m}{n}
\rput(1.4223437,3.062233){$x_2$}
\psline[linewidth=0.04cm,arrowsize=0.06cm 3.0,arrowlength=2.0,arrowinset=0.0]{->}(1.6009375,-2.5477672)(9.500937,-2.5477672)
\usefont{T1}{ptm}{m}{n}
\rput(9.122344,-2.937767){$x_1$}
\psline[linewidth=0.02cm,linestyle=dashed,dash=0.16cm 0.16cm](6.1209373,2.352233)(6.1209373,-2.647767)
\psline[linewidth=0.02cm,linestyle=dashed,dash=0.16cm 0.16cm](8.240937,0.7522329)(1.7209375,0.7522329)
\psline[linewidth=0.04cm,arrowsize=0.04cm 3.0,arrowlength=2.0,arrowinset=0.0]{->}(6.2009373,-1.4477671)(6.9009376,-2.147767)
\psline[linewidth=0.04cm,arrowsize=0.04cm 3.0,arrowlength=2.0,arrowinset=0.0]{->}(3.9009376,1.3522329)(3.2009375,2.052233)
\psline[linewidth=0.04cm,arrowsize=0.04cm 3.0,arrowlength=2.0,arrowinset=0.0]{->}(7.1009374,1.7522329)(7.8009377,2.4522328)
\usefont{T1}{ptm}{m}{n}
\rput(8.222343,2.4622328){$\mat{e}_1$}
\usefont{T1}{ptm}{m}{n}
\rput(2.8223438,2.062233){$\mat{e}_3$}
\usefont{T1}{ptm}{m}{n}
\rput(7.322344,-2.137767){$\mat{e}_2$}
\usefont{T1}{ptm}{m}{n}
\rput(4.092344,-3.0377672){$x^0_1$}
\usefont{T1}{ptm}{m}{n}
\rput(1.3923438,-0.7377671){$x^0_2$}
\psline[linewidth=0.02cm,linestyle=dashed,dash=0.16cm 0.16cm](6.1209373,2.392233)(1.7209375,2.392233)
\psline[linewidth=0.02cm,linestyle=dashed,dash=0.16cm 0.16cm](8.240937,0.7122329)(8.240937,-2.607767)
\usefont{T1}{ptm}{m}{n}
\rput(8.252344,-2.997767){$\snorm{\bh_1}$}
\usefont{T1}{ptm}{m}{n}
\rput(1.2323438,2.4622328){$\snorm{\bh_2}$}
\usefont{T1}{ptm}{m}{n}
\rput(5.02625,-0.00776709){\Large $\Omega$}
\usefont{T1}{ptm}{m}{n}
\rput(6.1923437,-3.0377672){$x^*_1$}
\usefont{T1}{ptm}{m}{n}
\rput(1.4723438,0.8622329){$x^*_2$}
\usefont{T1}{ptm}{m}{n}
\rput(1.7523438,-3.0377672){$0$}
\usefont{T1}{ptm}{m}{n}
\rput(1.3523438,-2.5377672){$0$}
\rput{-270.0}(8.33317,-3.6087046){\psarc[linewidth=0.04](5.9709377,2.362233){0.15}{-180.0}{0.0}}
\rput{-270.0}(8.63317,-3.9087045){\psarc[linewidth=0.04](6.2709374,2.362233){0.15}{0.0}{180.0}}
\psarc[linewidth=0.04](8.220938,0.6122329){0.14}{0.0}{180.0}
\rput{-180.0}(16.441875,1.7844658){\psarc[linewidth=0.04](8.220938,0.8922329){0.14}{0.0}{180.0}}
\rput{-270.0}(1.5931704,-6.7087045){\psarc[linewidth=0.04](4.1509376,-2.5577672){0.15}{0.0}{180.0}}
\rput{-180.0}(3.641875,-1.4155341){\psarc[linewidth=0.04](1.8209375,-0.70776707){0.14}{0.0}{180.0}}
\end{pspicture}
}
\caption{An illustration of a two-dimensional gain-region and its upper boundaries in directions $\mat{e}_1 = [1,1], \mat{e}_2 = [1,-1]$, and $\mat{e}_3 = [-1,1]$.}
\label{fig:gain}
\end{figure}

An illustration of a two-dimensional power gain-region is given in \figurename~\ref{fig:gain}. The direction vectors $\mat{e}_1, \mat{e}_2$, and $\mat{e}_3$ refer to three different parts of the boundary. For the choice $\mat{e} = \mat{1}$, the upper boundary in direction $\mat{e}$ is the usual upper boundary as in Definition \ref{def:uppBoundary}, i.e., $\partial^+ \Omega = \partial^{\mat{1}} \Omega$.

We draw an example to illustrate the importance of the boundary parts of the gain-region. Assume as in \figurename~\ref{fig:gain}, there exist two single antenna receivers and a transmitter. Assume, receiver $1$ is the intended receiver of the transmitter, i.e. $1=\overline{\mathcal{K}}$, and receiver $2$ is its unintended receiver such that $2=\underline{\mathcal{K}}$. For efficient operation in the setting, the transmitter is interested in maximizing its power gain at receiver $1$ and also interested in minimizing the power gain on receiver $2$. If we seek to characterize the set of efficient transmission strategies of the transmitter, then the boundary part corresponding to $\mat{e}_2 = [1,-1]$ is relevant. Any power gain tuple which is inside the power gain-region and does not lie on its boundary is not relevant, since the transmitter can increase the power gain to its intended receiver without changing the power gain to the unintended receiver. The direction vector $\mat{e}_2$ corresponds to the boundary part which includes the maximum achievable power gain at receiver 1 and also the minimum achievable power gain at receiver 2. These extreme points correspond to MRT and ZF transmission strategies, respectively. According to \figurename~\ref{fig:gain}, MRT achieves $\snorm{\bh_{1}}$ power gain at the first receiver and $x_2^*$ at the second receiver. On the other hand, ZF transmission achieves zero power gain at the second receiver and $x_1^0$ at the first receiver. From this example, it can be observed that the boundary part which corresponds to the transmitters efficient strategy set is the one where the direction vector has positive component corresponding to the intended receiver and negative component corresponding to the unintended receiver. Since a single transmitter has at least one intended receiver, otherwise it will not operate, the direction vector with all components equal to $-1$ is not of interest. Therefore, we define the set of feasible directions
\begin{equation}\label{def:direction_set}
\mathcal{E} := \{-1,1\}^K \setminus \{-1\}^K.
\end{equation}
\noindent Next, we study the transmission strategies that achieve the boundary points in $\partial^{\mat{e}} \Omega, \mat{e} \in \mathcal{E}$.

The relation between the number of existing receivers and the number of available antennas at the transmitter is important to distinguish whether power control is needed to achieve all boundary parts of the power gain-region. First, we consider the case where the number of transmit antennas is greater than or equal to the number of existing receivers $K$. For this case, a comprehensive study of the gain-region is given due to which we gain a link and some insights to a mathematical field of research in matrix analysis. Afterwards, we consider the case in which the number of antennas is strictly less than $K$. This case is addressed more briefly since the tools needed for the analysis are similar to those in the first case.
\subsection{The number of transmit antennas satisfies $N \geq K$}\label{sec:NgeqK}%
In this section, we assume $N \geq K$ and the channel vectors to the receivers are independently distributed. These assumptions lead to linear independence of the channel vectors with probability one. In this case, it is possible to achieve power gain on one receiver and simultaneously null the power gain at the remaining receivers. Therefore, the gain-region has boundary points that lie on each axis as the points $x_1^0$ and $x_2^0$ in \figurename~\ref{fig:gain}. As a result, all upper boundary points of the gain-region are achieved with full power. In case the number of antennas at the transmitter is strictly larger than $K$, the gain-region has part of its boundary the points between $x_1^0$ and the origin.
%
%
\begin{lemma}\label{thm:lemma_trace1}
Transmit covariance matrices from the set
\begin{equation}\label{eq:S_trace}
\widehat{\mathcal{S}} := \br{\bQ : \bQ \succeq 0, \tr{\bQ} = 1},
\end{equation}
achieve all points in $\partial^{\mat{e}} \Omega, \mat{e} \in \mathcal{E}$.
\end{lemma}
\begin{IEEEproof}
The proof is provided in Appendix \ref{proof:lemma_trace1}.
\end{IEEEproof}
%
%
Lemma \ref{thm:lemma_trace1} supports the result in \cite{Larsson2008}, where it is shown that all Pareto efficient operating points in the MISO IC correspond to full power transmission when the number of antennas is larger than or equal to the number of receivers. The power gain-region achieved with full power transmit covariance matrices is defined as
\begin{eqnarray}\label{eq:Omega_hat}
	\widehat{\Omega} := \left\{ \pp{x_{1}(\bQ),...,x_{K}(\bQ)} :  \bQ \in \widehat{\mathcal{S}} \right\} \subset \mathbb{R}_+^K.
\end{eqnarray}
\noindent Since the power gain at a receiver $\ell$ can equivalently be formulated as $x_{\ell}(\bQ) = \mat{h}_{\ell}^H \mat{Q} \mat{h}_{\ell} = \tr{\mat{Q} \mat{h}_{\ell} \mat{h}_{\ell}^H}$, the set $\widehat{\Omega}$ in \eqref{eq:Omega_hat} is rewritten as
\begin{equation}
\widehat{\Omega} = \br{\pp{\tr{\bQ \bh_1 \bh_1^\H},...,\tr{\bQ \bh_K \bh_K^\H}} :  \bQ \succeq 0, \tr{\bQ} = 1}.
\end{equation}
\noindent This set is referred to in \cite{Barker1984} as the \emph{joint field of values} of the set of matrices $\bh_1\bh_1^\H,...,\bh_K\bh_K^\H$. The set $\widehat{\Omega}$ is compact and convex\footnote{The set $\widehat{\Omega}$ is the convex hull of the \emph{joint numerical range} of the matrices $\bh_1\bh_1^\H,...,\bh_K\bh_K^\H$. The joint numerical range of a set of $K$ matrices $\mat{A}_1,...,\mat{A}_K$ is defined as \cite{Bonsall1973}: $\mathcal{W}\pp{\mat{A}_1,...,\mat{A}_K} := \br{\pp{\bz^\H \mat{A}_1 \bz,...,\bz^\H \mat{A}_K \bz}:\bz^\H \bz = 1}.$ In our case, $A_1=\bh_1\bh_1^\H,...,A_K=\bh_K\bh_K^\H$. Convexity of the set $\mathcal{W}\pp{\mat{A}_1,...,\mat{A}_K}$ is not always satisfied. Conditions for its convexity are studied in \cite{Li2000}.}. The next result shows that the boundary of $\widehat{\Omega}$ in any direction $\mat{e} \in \mathcal{E}$ can be achieved with rank-1 transmit covariance matrices. 
%
%
\begin{lemma}\label{thm:lemma_rank1}
Transmit covariance matrices from the set
\begin{equation}\label{eq:S}
\widetilde{\mathcal{S}} := \br{\bQ : \bQ \in \widehat{\mathcal{S}}, \rank{\bQ} = 1},
\end{equation}
achieve all points in $\partial^{\mat{e}} \widehat{\Omega}, \mat{e} \in \mathcal{E}$.
\end{lemma}
\begin{IEEEproof}
The proof is provided in Appendix \ref{proof:lemma_rank1}.
\end{IEEEproof}
Lemma \ref{thm:lemma_rank1} supports the result in \cite{Shang2009a} as a special case, where it is shown that single-stream beamforming is optimal in the MISO IC to achieve Pareto efficient operating points. Accordingly, efficient transmission strategies can be described by beamforming vectors.
~The next theorem characterizes the beamforming vectors that achieve the upper boundary of the set $\widehat{\Omega}$ in a specific direction $\mat{e}$.
%
%
\begin{theorem}
\label{theo:1}
All upper boundary points of the set $\widehat{\Omega}$ in direction $\mat{e} \in \mathcal{E}$ can be achieved by
	\begin{eqnarray}\label{eq:main}
		\bw(\mat{\lambda}) = \mat{v}_{max} \Biggl( \underbrace{\sum_{\ell=1}^K \lambda_\ell e_{\ell} \bh_{\ell} \bh_{\ell}^\H}_{\mat{Z}} \Biggr),
	\end{eqnarray}
	with
\begin{eqnarray}
	\mat{\lambda} \in \mat{\Lambda} := \left\{ \mat{\lambda} \in [0,1]^K : \sum_{\ell=1}^K \lambda_\ell = 1 \right\}. \label{eq:Lambda}
\end{eqnarray}
\end{theorem}

\begin{IEEEproof}
The proof is provided in Appendix \ref{proof:thm1}.
\end{IEEEproof}
\noindent The interesting observation from Theorem \ref{theo:1} is that all upper boundary points in direction $\mat{e}$ of the $K$-dimensional gain-region can be achieved by a parametrization using $K-1$ real parameters. Note that the components of the direction vector $\mat{e}$ in Theorem \ref{theo:1} can have positive and negative components. For the particular choices $\lambda_\ell = 0$ for all $\ell$ such that $e_\ell = +1$, the largest eigenvalue of $\mat{Z}$ in \eqref{eq:main} is zero. In this case the largest eigenvalue of $\mat{Z}$ can also have geometric multiplicity larger than one, i.e., there exist multiple linearly independent eigenvectors associated with that eigenvalue. For example, assume $N = 3, K = 2, e_1 = +1,$ and $e_2=-1$. For $\lambda_1 = 0$ and $\lambda_2 = 1$, the matrix $\mat{Z}$ in Theorem \ref{theo:1} is equal to $-\bh_2\bh_2^\H$ which is rank-1 with the largest eigenvalue having geometric multiplicity two. In this case, we choose an eigenvector which lies in the span of $\mat{H} = [\bh_1,...,\bh_K]$.

\begin{figure}[t]
\centering
\includegraphics[width=8.8cm]{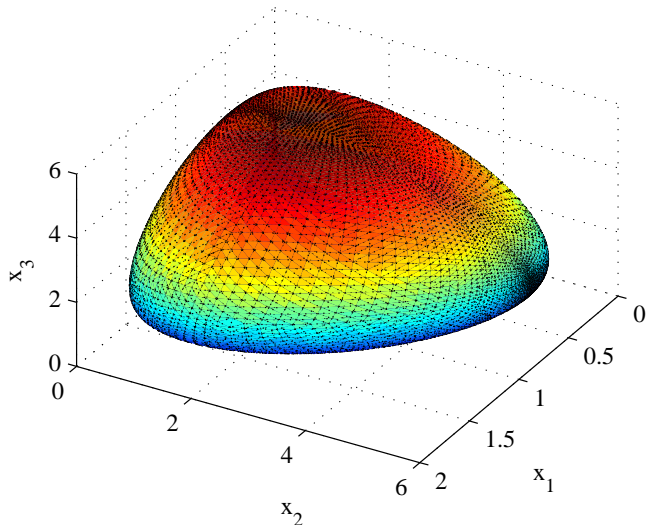}
\caption{An illustration of a three dimensional gain-region where the transmitter uses three antennas.}
\label{fig:Gain3D3N}
\end{figure}

In \figurename~\ref{fig:Gain3D3N}, a three dimensional gain-region is plotted where the transmitter uses three transmit antennas. The boundary points are calculated by generating the beamforming vectors characterized in Theorem \ref{theo:1}. For each boundary part, two real-valued parameters are required, and these are varied between zero and one with a step-length of $0.02$. The gain-region is shown to have a convex boundary. Moreover, since it is possible to null out the power gain at two receivers simultaneously, the boundary of the gain-region touches each axis in one point.
\subsection{The number of transmit antennas satisfies $N < K$}\label{sec:NlessK}%
In case the number of antennas at the transmitter satisfies $N < K$, it is not possible for the transmitter to choose a full power transmission strategy which nulls out the power gain at all except one receiver. Hence, the transmitter's ZF strategy would be to allocate no transmit power. This reveals that in order to achieve all boundary points of the power gain-region the transmitter has to reduce its transmission power. This is the reason why we study this case separately.

We start by assuming that the transmit covariance matrices are chosen from the set $\mathcal{S}$ in \eqref{eq:S_set}. The corresponding power gain-region is compact and convex according to Lemma \ref{thm:lemma_conv}. Hence the following programming problem, similar to the one formulated in \eqref{prb:origin} in Appendix \ref{proof:lemma_rank1}, achieves the boundary points of $\Omega$ in direction $\mat{e}\in\mathcal{E}$:
%
%
\begin{equation}\label{prb:progrank1}
\begin{split}
\text{maximize}& \quad \sum_{\ell = 1}^{K} \lambda_{\ell}e_\ell \bh_\ell^\H \bQ \bh_\ell\\
\text{subject to}& \quad \tr\bQ \leq 1, \quad \bQ \succeq 0,
\end{split}
\end{equation}
\noindent where $\mat{\lambda} \in \mat{\Lambda}$ is defined in \eqref{eq:Lambda}. As in Appendix \ref{proof:lemma_rank1}, the problem in \eqref{prb:progrank1} can be equivalently written as
\begin{equation}\label{prb:progrank1_eig}
\begin{split}
\text{maximize}& \quad \sum_{\ell = 1}^{N} \mu_\ell(\bQ) \mu_\ell \Biggl(\underbrace{\sum_{\ell = 1}^{K} \lambda_{\ell}e_\ell \bh_\ell \bh_\ell^\H}_{\mat{Z}}  \Biggr)\\
\text{subject to}& \quad \sum_{\ell=1}^{N} \mu_{\ell}(\bQ) \leq 1, \quad \mu_\ell(\bQ) \geq 0, \quad \text{for all } \ell = 1,...,N
\end{split}
\end{equation}
\noindent The solution of this problem is $\mu_{N}(\bQ) = p, p \in [0,1],$ and $\mu_{\ell}(\bQ) = 0$ for $\ell \neq N$. Hence, the optimal transmit covariance matrices $\bQ$ are rank-1. In addition, the optimal power allocation is to allocate power only in the direction of the eigenvector corresponding to the largest eigenvalue of $\mat{Z}$. Hence the formulation in Theorem \ref{theo:1} is still valid to determine the beamforming vectors that achieve the boundary points of $\Omega$. However, power control is to be applied on specific beamforming vectors. In order to maximize the problem in \eqref{prb:progrank1_eig}, we choose  $p = 1$ if the largest eigenvalue of $\mat{Z}$ is strictly larger than zero, and $p = 0$ if the largest eigenvalue of $\mat{Z}$ is strictly less than zero. It can be easily checked that $\mat{Z}$ can be negative definite according to Weyl's eigenvalue inequality in \cite[Theorem 4.3.7]{Horn1985}. We give a simple example to illustrate this case. Assume $N = 2, K = 3, e_1 = +1,$ and $e_2=e_3=-1$. For $\lambda_1 = 0$ and $\lambda_2,\lambda_3 > 0$, then $\mat{Z} = -\lambda_2\bh_2\bh_2^\H - \lambda_3\bh_3\bh_3^\H$ which is negative definite. In case the largest eigenvalue of $\mat{Z}$ is zero, all feasible power allocations should be adopted in order to achieve the boundary of the gain region. For the same example given above and for $\mu(\mat{Z})=0$, the weighted sum gains in \eqref{eq:mma} from Appendix \ref{proof:thm1} formulates to $\lambda_1 x_1(\bw_1\bw_1^\H) = \lambda_2 x_2(\bw_1\bw_1^\H) + \lambda_3 x_3(\bw_1\bw_1^\H)$. This equation, with the variables in $\mat{\lambda}$, describes the plane that touches the boundary of the power gain-region at $\mat{x}(\bw_1\bw_1^\H)$ and passes through the origin. Choosing a power allocation $p \in [0,1]$ achieves points on the segment connecting the origin and $\mat{x}(\bw_1\bw_1^\H)$ which correspond to boundary points on the power gain-region. The power allocations that achieve the boundary points of the power gain-region $\Omega$ in direction $\mat{e} \in \mathcal{E}$ are summarized as follows:
\begin{eqnarray}\label{cases:NlessK}
	p = 1 \text{ for } \mu_N\pp{\mat{Z}} > 0, \quad p \in [0,1] \text{ for } \mu_N \pp{\mat{Z}} = 0, \quad p = 0 \text{ for } \mu_N\pp{\mat{Z}} < 0.
\end{eqnarray}
\noindent where $\mat{Z}$ is given in \eqref{prb:progrank1_eig}.

\begin{figure}[t]
\centering
\includegraphics[width=8.8cm]{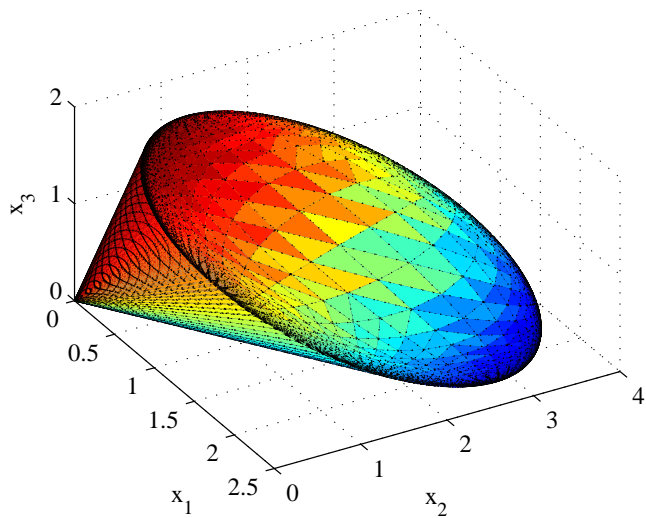}
\caption{An illustration of a three dimensional gain-region where the transmitter uses two antennas.}
\label{fig:Gain3D}
\end{figure}

In \figurename~\ref{fig:Gain3D}, a three dimensional gain-region is plotted where two antennas are utilized at the transmitter. In comparison to the plot in \figurename~\ref{fig:Gain3D3N}, the region looks like a cone whose vertex is at the origin. The outermost boundary which is not flat is attained with full power transmission. The flat surfaces correspond to the case for which the transmission power is varied between zero and one.

\section{Pareto Boundary Characterization}\label{sec:pareto}%
In this section, all $T$ transmitters are considered again. The analysis of the single transmitter case in the previous section builds the framework to determine the beamforming vectors for each transmitter that are relevant for Pareto optimal operation. Each transmitter $k,k \in \mathcal{T},$ is associated with a power gain-region $\Omega_k$. The sets of intended and unintended receivers of a transmitter $k$ are $\overline{\mathcal{K}}(k)$ and $\underline{\mathcal{K}}(k)$, respectively. In a general network, efficient operation of the systems requires the transmitters to maximize the power gain at intended receivers and simultaneously minimize the power gain at unintended receivers. In order to achieve Pareto optimal operating points a combination of these objectives is required at each transmitter. The gain-region characterization in the previous section, illustrates the effects of the beamforming vectors on the power gains achieved jointly at the receivers. Next, we formalize the transmit beamforming vectors of each transmitter that are relevant to achieve Pareto optimal points in the utility region.

\begin{theorem}\label{thm:thm2} All Pareto optimal points in the utility region $\mathcal{U}$ can be achieved by beamforming vectors
	\begin{eqnarray}\label{eq:BF_pareto}
		\mat{w}_k(\mat{\lambda}_k) = p_k \mat{v}_{max} \Biggl(\underbrace{\sum_{\ell=1}^K \lambda_{k,\ell} e_{k,\ell} \mat{h}_{k,\ell} \mat{h}_{k,\ell}^H}_{\mat{Z}_k}\Biggr),  \label{eq:resmisoifc}
	\end{eqnarray}
	with $\mat{\lambda}_k \in \mat{\Lambda}$ defined in (\ref{eq:Lambda}) and
	\begin{eqnarray}\label{cases:e}
		e_{k,\ell} = \begin{cases} +1 & \ell \in \mathcal{\overline{K}}(k) \\
		-1 & \ell \in \mathcal{\underline{K}}(k) \end{cases},
	\end{eqnarray}
\begin{eqnarray}\label{cases:NlessKthm}
	p_k = \begin{cases} 1 & \mu_N\pp{\mat{Z}_k} > 0\\
    [0,1] & \mu_N\pp{\mat{Z}_k} = 0\\
    0 & \mu_N\pp{\mat{Z}_k} < 0 \end{cases}.
\end{eqnarray}

	\label{theo:3}
\end{theorem}

\begin{IEEEproof}
The proof is provided in Appendix \ref{proof:thm2}.
\end{IEEEproof}

Theorem \ref{thm:thm2} characterizes the transmission strategies that achieve Pareto optimal points. The number of real-valued parameters, $\mat{\lambda}$ and $p$, that are required to characterize these beamforming vectors is: (i) $T(K-1)$ in case no power control is needed. This correspond to the case when $N_k > \abs{\underline{\mathcal{K}}(k)}$. (ii) $TK$ in case power control is needed corresponding to the case $N_k \leq \abs{\underline{\mathcal{K}}(k)}$. All these parameters take values between zero and one. The direction vector in \eqref{cases:e} can be determined since each transmitter $k$ knows its intended and unintended receiver sets, $\mathcal{\overline{K}}(k)$ and $\mathcal{\underline{K}}(k)$ respectively. This direction vector specifies the relevant boundary of the transmitter's power gain-region.

\subsection{Example Revisited}\label{sec:example}%
\begin{figure}[t]
\centering
\includegraphics[width=8.8cm]{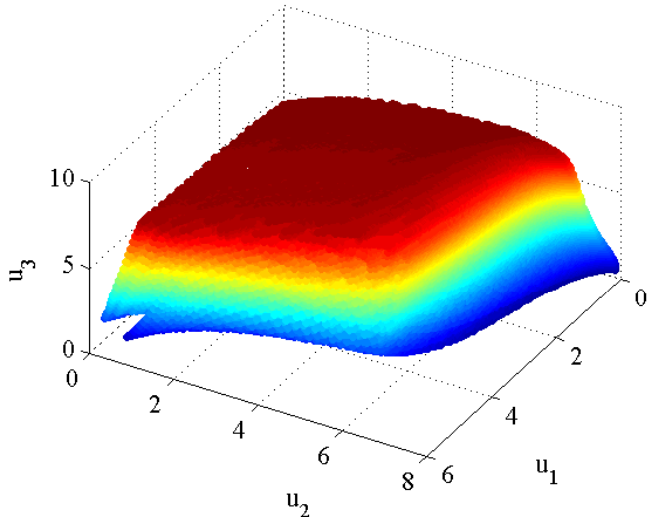}
\caption{Pareto boundary of the utility region of the setting described in Section \ref{sec:example} with SNR = 15 dB and $N=3$.}
\label{fig:Pareto3D}
\end{figure}

The example in Section \ref{sec:example_setting} will be continued to find the beamforming vectors that achieve Pareto optimal points in the corresponding utility region. The choice of the beamforming vectors with unit norm for transmitters $11$, $12$ and $2$ which achieve Pareto optimal points are characterized in Theorem \ref{thm:thm2}. Since the number of antennas at each transmitter is strictly larger than the number of unintended receivers, all transmitters should transmit at full power according to \eqref{cases:NlessKthm}. The power allocation $p_{11}$ and $p_{12}$ on the beamforming vectors $\bw_{11}$ and $\bw_{12}$, respectively, is varied such $p_{11} + p_{12} = 1$. The power allocation parameters $p_{11}$ and $p_{12}$ can be expressed as $q$ and $1-q$, respectively, with $q \in [0,1]$. The characterization in Theorem \ref{thm:thm2} leads to the following nonnegative real-valued parameters:
\begin{itemize}
\item For transmitter $11$: $\lambda_{11,1}, \lambda_{11,2}, \lambda_{11,3}$, with $\lambda_{11,1} + \lambda_{11,2} + \lambda_{11,3} = 1$.

\item For transmitter $12$: $\lambda_{12,1}, \lambda_{12,2}, \lambda_{12,3}$, with $\lambda_{12,1} + \lambda_{12,2} + \lambda_{12,3} = 1$.

\item For transmitters $11$ and $12$: $q \in [0,1]$.

\item For transmitter $2$: $\lambda_{2,1}, \lambda_{2,2}, \lambda_{2,3}$, with $\lambda_{2,1} + \lambda_{2,2} + \lambda_{2,3} = 1$.
\end{itemize}

\noindent All seven required parameters are in the interval $[0,1]$, and the plot in \figurename~\ref{fig:Pareto3D} is obtained by varying these in a grid with $0.05$ step-length. The points obtained include the points that lie on the Pareto boundary of the utility region which satisfy Definition \ref{def:1}. Points corresponding to weak Pareto optimality are not achieved by the parametrization in Theorem \ref{thm:thm2} and thus not included in the plot. Weak Pareto optimality is defined as in Definition \ref{def:1} except that the corresponding inequality is not strict. Weak Pareto optimal points complete the shape of the utility region by orthogonally projecting each Pareto optimal point in \figurename~\ref{fig:Pareto3D} onto the coordinate surfaces.

In the next section, we discuss two special applications of the developed framework.

\section{Applications}\label{sec:applications}

\subsection{Multiple-input Single-output Interference Channel}\label{sec:applications1}
The $K$-users MISO IC consists of $K$ transmitter-receiver pairs, where each receiver has an intended transmitter while all other transmitters induce interference on this receiver. We consider single-user decoding, i.e., interference is treated as additive noise at each receiver. For a given set of beamforming vectors $\{\mat{w}_1,...,\mat{w}_K\}$, the following rate is achievable at receiver $k$, by using codebooks approaching Gaussian ones:%
\begin{equation}\label{eq:r1}
\begin{split}
  u_k(\mat{w}_1\mat{w}_1^\H,...,\mat{w}_K\mat{w}_K^\H) =\log_2 \pp{1+\frac{\sabs{\mat{w}_k^\H \mat{h}_{kk}}}{\sum_{\ell\neq k} \sabs{\mat{w}_{\ell}^\H\mat{h}_{\ell k}}+\sigma^2}}.
\end{split}
\end{equation}
\noindent This utility function satisfies properties A and B in Assumption \ref{def:utility} which leads to the following receiver sets for each transmitter $k$: $\overline{\mathcal{K}}(k) = \br{k}$, and $\underline{\mathcal{K}}(k) = \mathcal{K} \backslash \{k\}$. All points on the Pareto boundary of the achievable rate-region of the MISO IC can be reached by beamforming vectors as given in Theorem \ref{thm:thm2}. In \cite{Jorswieck2008}, a characterization of the beamforming vectors that achieve the Pareto boundary of the achievable rate-region is provided by a complex linear combination of the MRT and ZF strategies. The parametrization in Theorem \ref{thm:thm2} is real-valued with the same number of parameters, thus of lower dimension.

\begin{figure}[t]
\begin{center}
\includegraphics[width=8cm]{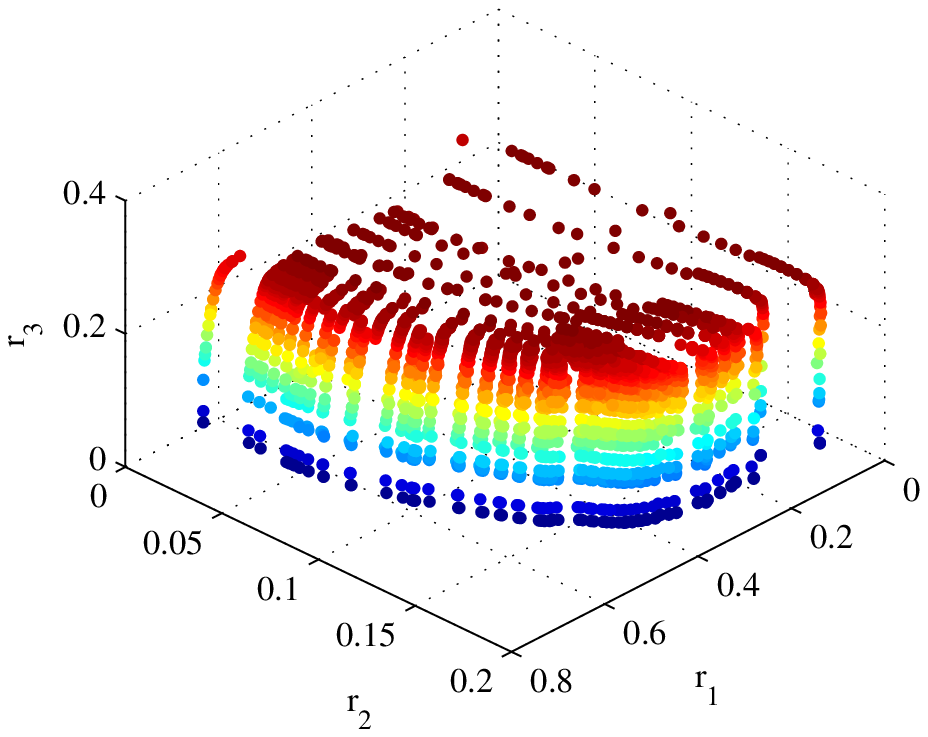}
\includegraphics[width=8cm]{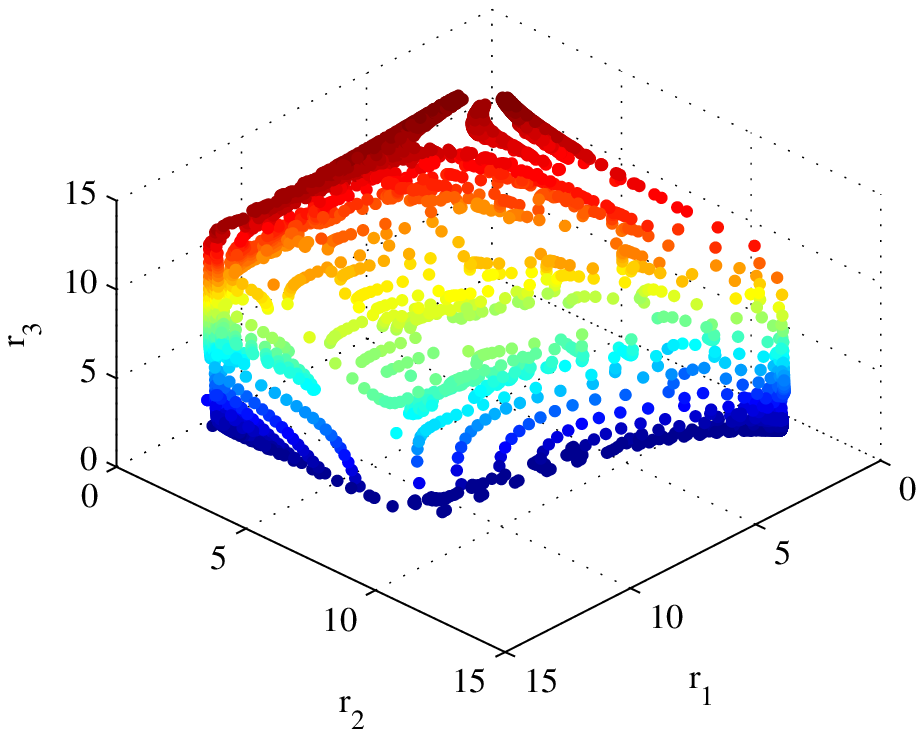}
\end{center}
\caption{Pareto boundary of a three user MISO IC rate-region, where $N = 3$. In the figure on the left, SNR = -10 dB, and in the figure on the right, SNR = 30 dB.}
\label{fig:RR3}
\end{figure}

The Pareto boundary of a three user MISO IC rate-region is plotted in \figurename~\ref{fig:RR3} (left) for SNR = $-10$ dB and in \figurename~\ref{fig:RR3} (right) for SNR = $30$ dB. The generated points correspond to the beamforming vectors characterized in Theorem \ref{thm:thm2}. The real-valued parameters are varied in a $0.05$ step-length. Since we are only interested in revealing the Pareto boundary of the achievable rate region, we do the following. We randomly choose ten thousand generated points and remove all points that are dominated by these. In other words, for a randomly chosen rate tuple, all points corresponding to joint rates less than the chosen one are removed. In addition, the algorithm provided in \cite{Katz2007} is applied. The algorithm reduces the number of plotted points in removing the points that are not visible from the viewed angle of the figure. This algorithm further reduces the complexity of rendering the generated points.

Next, we identify special operation points of each transmitter.

\subsubsection{Maximum ratio transmission}

Maximum ratio transmission for transmitter $k$ corresponds obviously to the parameters $\lambda_{k,\ell} = 1$ for $k = \ell$ and $\lambda_{k,\ell} = 0$ for $k \neq \ell$. This transmission strategy is the unique Nash equilibrium (NE) strategy of each transmitter $k$, which is the outcome of a noncooperative game between the users.

\subsubsection{Zero-forcing transmission}
Zero-forcing transmission is characterized with the following lemma:
\begin{lemma}\label{thm:zf}
 Zero Forcing transmission for transmitter $k$ corresponds to the parameters $\lambda_{k,\ell} = 0$ for $k = \ell$ and $\lambda_{k,\ell} > 0$ for $k \neq \ell$.
\end{lemma}

\begin{IEEEproof}
The proof is provided in Appendix \ref{proof:zf}.
\end{IEEEproof}

The two-user MISO IC special case has an appealing form in terms of the parametrization of the Pareto boundary of the achievable rate-region. The efficient beamforming vectors are a linear combination of the MRT and ZF strategies \cite[Corollary 2]{Jorswieck2008}. These two strategies have the interpretation of a transmitter being either selfish or altruistic \cite{Jorswieck2008b,Larsson2008}. The efficient beamforming vectors of transmitter $k \in \br{1,2}$ are given as \cite[Corollary 2]{Jorswieck2008}%
\begin{equation}\label{eq:NEZFcomb}
\bw_k (\hat{\lambda}_k) = \frac{\hat{\lambda}_k \bw_k^{\text{MRT}} + (1 - \hat{\lambda}_k) \bw_k^{\text{ZF}}}{\norm{\hat{\lambda}_k \bw_k^{\text{MRT}} + (1 - \hat{\lambda}_k) \bw_k^{\text{ZF}}}},
\end{equation}
\noindent where $\hat{\lambda}_k \in [0,1]$, $\bw_k^{\text{MRT}} = \frac{\bh_{kk}}{\norm{\bh_{kk}}},$ and $\bw_k^{\text{ZF}} = \frac{\mat{\Pi^\perp_{\bh_{k \ell}}} \bh_{kk}}{\norm{\mat{\Pi^\perp_{\bh_{k \ell}}} \bh_{kk}}}, k \neq \ell.$ Next, we prove that the parametrization in \eqref{eq:NEZFcomb} has the same set of strategies as in Theorem \ref{thm:thm2} for $K = 2$. For this case, the eigenvalue equation for the hermitian matrix in Theorem \ref{theo:1} is written as $\pp{\lambda_1 \bh_{11} \bh_{11}^\H  - (1-\lambda_1)\bh_{12} \bh_{12}^\H} \bw_1 = \mu \bw_1$. This equation can be equivalently formulated to $\lambda_1 \snorm{\bh_{11}} \frac{\bh_{11} \bh_{11}^\H}{\snorm{\bh_{11}}}\bw_1  - (1-\lambda_1)\snorm{\bh_{12}} \frac{\bh_{12} \bh_{12}^\H}{\snorm{\bh_{12}}} \bw_1 = \mu \bw_1.$ Adding $(1- \lambda_1)\snorm{\bh_{12}}\bw_1$ on both sides of the equation leads to
\begin{equation}\label{eq:2usereig2}
\pp{\lambda_1 \snorm{\bh_{11}} \Pi_{\bh_{11}} + (1-\lambda_1)\snorm{\bh_{12}} \Pi^\perp_{\bh_{12}}}\bw_1 = (\mu + (1-\lambda_1)\snorm{\bh_{12}})\bw_1.
\end{equation}
\noindent The LHS of the equation states that the principal eigenvector is a linear combination of its orthogonal projection on $\bh_{11}$ and the orthogonal projection onto the orthogonal complement of $\bh_{12}$. Since the largest eigenvalue $\mu$ is larger or equal to zero, the weight in the RHS of \eqref{eq:2usereig2} is always positive. Hence, the optimal set of beamforming vectors can be equivalently characterized by \eqref{eq:NEZFcomb}.

\subsection{Noncooperative Underlay Cognitive Radio}\label{sec:applications2}

In an underlay cognitive radio scenario, secondary users can share the communication resources with primary users under the condition of not imposing quality of service (QoS) degradation to the primary systems. A limited QoS degradation to the primary users is described by interference temperature constraints (ITC) \cite{Haykin2005}. When no interference on the primary users is allowed, the constraint is said to be a null-shaping constraint. Motivated by the concept of null-shaping constraints \cite{Scutari2009}, the next result gives an alternative characterization of efficient transmission strategies. In order to be able to fulfill the null-shaping constraints, the number of applied antennas at the transmitter has to be greater than or equal to the number of primary receivers. We assume there exists virtual single-antenna primary receivers, and consider the efficient design of the null-shaping constraints. The following result has been presented in \cite{Jorswieck2010} without proof.

\begin{corollary}\label{thm:cor1}Assume $N_k \geq K$ and define the matrix
\begin{equation}\label{eq:nullConstraints}
Z_k(\mat{\lambda}_k) = \left[\mat{z}_1(\mat{\lambda}_k),..., \mat{z}_{\abs{\mathcal{\underline{K}}(k)}}(\mat{\lambda}_k), \mat{z}_{N_k - \abs{\mathcal{\overline{K}}(k)} +1}(\mat{\lambda}_k),...,\mat{z}_{N_k-1}(\mat{\lambda}_k)\right],
\end{equation}
\noindent where
\begin{equation}
\mat{z}_i(\mat{\lambda}_k) = \mat{v}_i \pp{\sum_{\ell=1}^K \lambda_{k,\ell} e_{k,\ell} \mat{h}_{k\ell} \mat{h}_{k\ell}^H},
\end{equation}
with $\mat{\lambda}_k \in \mat{\Lambda}$ defined in (\ref{eq:Lambda}) and $e_{k,\ell}$ defined in \eqref{cases:e}. All points on the Pareto boundary of the utility region $\mathcal{U}$ can be reached by beamforming vectors
\begin{equation}\label{eq:BF_null}
\bw_k(\mat{\lambda}_k) = \frac{ \mat{\Pi}_{Z_k(\mat{\lambda}_k)}^{\perp} \bh_{k\ell}}{\norm{\mat{\Pi}_{Z_k(\mat{\lambda}_k)}^{\perp}\bh_{k\ell}}}, \quad \ell \in \overline{\mathcal{K}}(k).
\end{equation}
\end{corollary}
\begin{IEEEproof}
The proof is provided in Appendix \ref{proof:null_NE}.
\end{IEEEproof}

In Corollary \ref{thm:cor1}, the design of the null-shaping constraints is given in \eqref{eq:nullConstraints}, and the efficient transmission strategies are given in \eqref{eq:BF_null}. Here, $K-1$ null-shaping constraints are to be applied on each transmitter, and  the number of required real-valued parameters is the same as in Theorem \ref{thm:thm2}. Hence, the complexity of parameterizing the efficient beamforming vectors is the similar in Corollary \ref{thm:cor1} and Theorem \ref{thm:thm2}.

The form of the transmission strategy in \eqref{eq:BF_null} has a relevant interpretation. Consider a MISO IC setting as in the previous section. Rewriting \eqref{eq:BF_null} for transmitter $k$ gives $\bw_k(\mat{\lambda}_k) = \frac{ \mat{\Pi}_{Z_k(\mat{\lambda}_k)}^{\perp} \bh_{kk}}{\norm{\mat{\Pi}_{Z_k(\mat{\lambda}_k)}^{\perp}\bh_{kk}}}.$ This transmission strategy is MRT that satisfies null-shaping constraints given in $Z_k(\mat{\lambda}_k)$. The MRT strategy is the unique noncooperative strategy of transmitter $k$, and corresponds to the NE strategy in game theoretical terms \cite{Osborne1994}. Through the design of the null-shaping constraints in Corollary \ref{thm:cor1}, all Pareto optimal points of the utility region are characterized by transmission strategies that are in NE and satisfy the characterized null-shaping constraints. The interesting observations are as follows. Null-shaping constraints are sufficient to characterize the Pareto boundary of the MISO IC rate region. Moreover, given the null-shaping constraints, the transmitters are required to be noncooperative in order to achieve efficient operating points \cite{Jorswieck2010,Jorswieck2010b}. This result can be exploited by clever secondary user selection algorithms.

\section{Extension to Multiple Antennas at the Receivers}\label{sec:MIMO}
We discuss the extension to multiple-antennas at the receivers and study which settings can be applied to our framework. Assume that the number of antennas used by transmitter $k$ is $N_k$ and the number of receive antennas at receiver $\ell$ is $R_\ell$. The channel matrix from transmitter $k$ to receiver $\ell$ is denoted by $\mat{H}_{k \ell} \in \mathbb{C}^{R_k \times N_k}$. The signal after receive filtering at receiver $\ell$ is
\begin{eqnarray}
y_\ell= \sum_{k=1}^T \mat{z}_\ell^\H \mat{H}_{k \ell} \mat{w}_k s_k + \mat{z}_\ell^\H \mat{n}_\ell, \label{eq:systemmodelMIMO}
\end{eqnarray}
\noindent where $\mat{w}_k$ and $\mat{z}_\ell$ are transmit and receive beamforming vectors, respectively. Considering a single transmitter $k$ as in Section \ref{sec:powergain}, the power gain achieved at receiver $\ell$ is $x_{k\ell}(\mat{w}_k,\mat{z}_\ell) = \sabs{\mat{z}_\ell^\H \mat{H}_{k \ell} \mat{w}_k}$. Clearly, if the receive beamforming vectors do not depend on the transmit beamforming vector $\bw_k$, then the framework in this paper can be applied to the setting. For example, the receive beamforming vectors $\mat{z}_\ell$ can be fixed prior to transmission as in singular value decomposition (SVD) based receive beamforming. Let the SVD\footnote{$\mat{H} = \mat{U}_{\mat{H}} \mat{\Sigma}_{\mat{H}} \mat{V}_{\mat{H}}^\H$ where $\mat{U}_{\mat{H}}$ is the $R_k \times R_k $ unitary matrix having the left singular vectors, $\mat{V}_{\mat{H}}$ is the $N_k \times N_k $ unitary matrix having the right singular vectors, and $\mat{\Sigma}_{\mat{H}}$ is the $R_k \times N_k $ diagonal matrix containing the nonnegative singular values.} of $\mat{H}_{k \ell}$ be $\mat{U}_{\mat{H}_{k \ell}} \mat{\Sigma}_{\mat{H}_{k \ell}} \mat{V}_{\mat{H}_{k \ell}}^\H$. The SVD based beamforming vector $\mat{z}_\ell^{\text{svd}}$ depends only on $\mat{H}_{k \ell}$ and is the left singular vector in $\mat{U}_{\mat{H}_{k \ell}}$ which corresponds to the largest singular value in $\mat{\Sigma}_{\mat{H}_{k \ell}}$. The framework in this paper can also be applied when the receive beamforming vectors depend only on intended transmitters' beamforming vectors. Assume the intended transmitter of receiver $\ell$ is $k$, then the maximum ration combining (MRC) beamforming vector of receiver $\ell$ is $\mat{z}_\ell^{\text{mrc}}(\bw_k) = \mat{H}_{k \ell} w_k$. In this case, the transmit beamforming vectors of the transmitter are not coupled through the receive beamforming vectors. Hence, the power gain region of a single transmitter can be used as a tool to characterize its efficient beamforming vectors. In case one receiver has multiple intended transmitters as receiver $2$ in the example in Section \ref{sec:example_setting}. Receiver $2$ uses $\mat{z}_{2,12} = \mat{H}_{(12)2}\mat{w}_{12}$ to successively decode the signal from transmitter $12$ and uses $\mat{z}_{2,2} = \mat{H}_{22}\mat{w}_{2}$ to successively decode the signal from transmitter $2$. The utility of receiver $2$ is $\log_2({1 + \sabs{\mat{w}_{12}^\H \mat{H}_{12}^\H \mat{H}_{12}\mat{w}_{12}} + \sabs{\mat{w}_{2}^\H \mat{H}_{22}^\H \mat{H}_{22}\mat{w}_{2}}})$, where the power gains from the transmitters are not coupled by the transmit beamforming vectors.

The framework in this paper will not apply if the receive beamforming vectors depend on several transmit beamforming vectors. An example is the linear minimum mean square error (MMSE) receiver \cite{Viswanath1999}. Assume the intended transmitter of receiver $\ell$ is $k$, then the linear MMSE receiver is $z_\ell^{\text{mmse}}(\bw_1,\ldots,\bw_T) = (\sigma^2 \mat{I} + \sum_{j \neq k} \mat{H}_{j \ell} \mat{w}_j \mat{w}_j^\H \mat{H}_{j \ell}^\H)^{-1} \mat{H}_{k \ell} \mat{w}_k$. Thus the transmit beamforming vectors $\bw_1,\ldots,\bw_T$ are coupled through the receive beamforming vector. Hence, the power gain-regions of the transmitters are coupled by their beamforming vectors.

\section{Conclusions}
In this work, we consider MISO wireless interference networks in which $T$ transmitters and $K$ receivers share the same spectral band. We provide a framework to determine the transmission strategies that are relevant for Pareto optimal operation. By studying the single transmitter's power gain-region, the properties of efficient transmission are acquired. We prove that the boundary of the gain-region is convex and achieved with single-stream beamforming. Due to the convexity of the boundary of the gain-region, the efficient transmit beamforming vectors can be parameterized by real-valued parameters. We determine and distinguish the conditions under which power control is required. When the number of antennas at the transmitter is greater than or equal to the number of existing receivers, we show that full power transmission achieves all boundary points of the gain region. In this case, the parameterizations of efficient beamforming vectors requires $K-1$ real-valued parameters between zero and one. When the number of antennas at the transmitter is strictly less than the number of receivers, we characterize the transmission strategies for which power control is required. For this case, an additional real-valued parameter is needed that varies the power level at the transmitter. We apply the single-transmitter framework to the multiple-transmitter case. On determining the important boundary part of each transmitter's gain-region, all Pareto efficient beamforming vectors are characterized. This parameterizations simplifies the design of Pareto efficient resource allocation schemes.

\section*{Acknowledgment}
The authors would like to thank Yiu Tung Poon, Erik Larsson, Christian Scheunert, Johannes Lindblom, and Eleftherios Karipidis for interesting discussions.

\appendices
\section{Proof of Lemma \ref{thm:lemma_conv}}\label{proof:lemma_conv}
Note that $\Omega$ does not correspond to the joint field of values \cite{Barker1984} which is known to be convex and compact. Therefore, we provide a proof of the convexity of $\Omega$ and show that it is bounded and closed, thus compact. It is simple to show that $\Omega$ is bounded because the power gain at the $\ell$th receiver has a finite maximum which is achieved when the transmitter performs MRT to that receiver, i.e., $x_{\ell}\pp{\bQ} \leq x_{\ell}\pp{\frac{\bh_{\ell} \bh_{\ell}^\H}{\bh_{\ell}^\H \bh_{\ell}}} = \snorm{\bh_\ell}.$ Therefore, the box described by the set $\mathcal{Y} := \br{\bx : 0 \leq x_\ell \leq \snorm{\bh_\ell}},$ contains $\Omega$, i.e., $\Omega \subset \mathcal{Y}$ as illustrated in \figurename~\ref{fig:gain}. The set $\Omega$ is closed because the feasible set of transmission strategies, $\bQ \succeq 0$ and $\tr{\bQ} \leq 1$, is compact and convex. Since every pre-image of a closed set is closed for continuous functions \cite{Rudin2002}, follows that $\Omega$ is a closed set.

It remains to prove that $\Omega$ is convex. For any two points $\bx(\bQ_x) \in \Omega$ and $\bx(\bQ_y) \in \Omega$, we prove that $\bx(\bQ_z) \in \Omega$, where $\bx(\bQ_z) = t\bx(\bQ_x) + (1-t)\bx(\bQ_y)$ and $t \in [0,1]$. Any component of $\bx(\bQ_z)$ is
\begin{equation}
x_{\ell}(\bQ_z) = t x_{\ell}(\bQ_x) + (1-t) x_{\ell}(\bQ_y) = t \bh_\ell^H \bQ_x \bh_\ell + (1-t) \bh_\ell^H \bQ_y \bh_\ell = \bh_\ell^H \pp{ t \bQ_x + (1-t) \bQ_y } \bh_\ell.
\end{equation}
\noindent Hence, the transmit covariance matrices that achieve the line segment between $\bx(\bQ_x)$ and $\bx(\bQ_y)$ are given as $\bQ_z (t) = t\bQ_x + (1-t)\bQ_y.$ Accordingly, since $\bQ_x$ and $\bQ_y$ are positive semidefinite, then $\bQ_z (t)$ is positive semidefinite for all $t \in [0,1]$. In addition, $\bQ_z (t)$ fulfills the trace constraint, $\tr{\bQ_z (t)} \leq 1$, for all $t \in [0,1]$ since $\tr{\bQ_z} = \tr{t\bQ_x + (1-t)\bQ_y} = t\tr{\bQ_x} + (1-t)\tr{\bQ_y} \leq 1.$ Therefore, $\bx(\bQ_z(t))$ also lies in $\Omega$ for all $t \in [0,1]$. Hence, the set $\Omega$ is a compact and convex set.

\section{Proof of Lemma \ref{thm:lemma_trace1}}\label{proof:lemma_trace1}
In order to prove that the transmit covariance matrices from $\widehat{\mathcal{S}}$ in \eqref{eq:S_trace} achieve points on the boundary of the set $\Omega$ in direction $\mat{e}$, $\mat{e} \in \mathcal{E}$, we show that for any transmit covariance matrix $\mat{P}$ with $\tr{\mat{P}} < 1$, a transmit covariance matrix $\bQ$ from $\widehat{\mathcal{S}}$ can be constructed in which $\bx(\bQ)$ dominates $\bx(\mat{P})$ in direction $\mat{e}$ according to Definition \ref{def:vect_dominates}. Assume $e_{\ell} = +1$, we can construct $\bQ$ as $\bQ = \mat{P} + \pp{1 - \tr{\mat{P}}} \frac{\mat{\Pi}_{\mat{Z}}^{\bot} \bh_{\ell} \bh_{\ell}^\H \mat{\Pi}_{\mat{Z}}^{\bot}} {\norm{\mat{\Pi}_{\mat{Z}}^{\bot} \bh_{\ell} \bh_{\ell}^\H \mat{\Pi}_{\mat{Z}}}},$ where $\mat{Z} = \left[ \bh_1,...,\bh_{\ell-1},\bh_{\ell+1},...,\bh_K \right]$. Clearly, $\bQ$ is in the set $\widehat{\mathcal{S}}$. Since $K \leq N$, the dimension of the null space of $\mat{Z}$ is greater or equal to one, therefore the projection $\mat{\Pi}_{\mat{Z}}^{\bot}$ is not equal to the zero vector. The power gain achieved with $\bQ$ at the $k$th receiver, $k \in \mathcal{K}$, is%
\begin{equation}
x_k(\bQ) = \bh_k^\H\pp{\mat{P} + \pp{1 - \tr{\mat{P}}} \frac{\mat{\Pi}_{\mat{Z}}^{\bot} \bh_{\ell} \bh_{\ell}^\H \mat{\Pi}_{\mat{Z}}^{\bot}} {\norm{\mat{\Pi}_{\mat{Z}}^{\bot} \bh_{\ell} \bh_{\ell}^\H \mat{\Pi}_{\mat{Z}}}}}\bh_k.
\end{equation}
\noindent We distinguish two cases. For $k \neq \ell$, the power gain at the $k$th receiver is
\begin{equation}
\begin{split}
x_k(\bQ) &= \bh_k^\H\pp{\mat{P} + \pp{1 - \tr{\mat{P}}} \frac{\mat{\Pi}_{\mat{Z}}^{\bot} \bh_{\ell} \bh_{\ell}^\H \mat{\Pi}_{\mat{Z}}^{\bot}} {\norm{\mat{\Pi}_{\mat{Z}}^{\bot} \bh_{\ell} \bh_{\ell}^\H \mat{\Pi}_{\mat{Z}}}}}\bh_k \\
&= \bh_k^\H\mat{P}\bh_k + \pp{1 - \tr{\mat{P}}}\underbrace{ \bh_k^\H\frac{\mat{\Pi}_{\mat{Z}}^{\bot} \bh_{\ell} \bh_{\ell}^\H \mat{\Pi}_{\mat{Z}}^{\bot}} {\norm{\mat{\Pi}_{\mat{Z}}^{\bot} \bh_{\ell} \bh_{\ell}^\H \mat{\Pi}_{\mat{Z}}}} \bh_k}_{=\mat{0}} = x_k(\mat{P}).
\end{split}
\end{equation}%
\noindent This implies that the gain at receiver $k \neq \ell$ has not changed. For $k = \ell$, the power gain at receiver $\ell$ is
\begin{equation}
\begin{split}
x_\ell(\bQ) &= \bh_\ell^\H\pp{\mat{P} + \pp{1 - \tr{\mat{P}}} \frac{\mat{\Pi}_{\mat{Z}}^{\bot} \bh_{\ell} \bh_{\ell}^\H \mat{\Pi}_{\mat{Z}}^{\bot}} {\norm{\mat{\Pi}_{\mat{Z}}^{\bot} \bh_{\ell} \bh_{\ell}^\H \mat{\Pi}_{\mat{Z}}}}}\bh_\ell
= \bh_\ell^\H\mat{P}\bh_\ell + \pp{1 - \tr{\mat{P}}}\bh_\ell^\H \frac{\mat{\Pi}_{\mat{Z}}^{\bot} \bh_{\ell} \bh_{\ell}^\H \mat{\Pi}_{\mat{Z}}^{\bot}} {\norm{\mat{\Pi}_{\mat{Z}}^{\bot} \bh_{\ell} \bh_{\ell}^\H \mat{\Pi}_{\mat{Z}}}}\bh_\ell \\
&= \bh_\ell^\H \mat{P}\bh_\ell + \pp{1 - \tr{\mat{P}}} \frac{\sabs{\bh_\ell^\H\mat{\Pi}_{\mat{Z}}^{\bot}\bh_\ell}}
{\norm{\mat{\Pi}_{\mat{Z}}^{\bot} \bh_{\ell} \bh_{\ell}^\H \mat{\Pi}_{\mat{Z}}}}
> x_\ell(\mat{P}).
\end{split}
\end{equation}
According to the above results we can construct $\bQ, \bQ \in \widehat{\mathcal{S}},$ such that $\bx(\bQ) \geq^{\mat{e}}\bx(\mat{P})$ for any given $\mat{P}$, with $\tr{\mat{P}} < 1$. Therefore, the boundary set $\partial^{\mat{e}} \Omega, \mat{e} \in \mathcal{E},$ can be achieved with transmit covariance matrices that fulfill the total power constraint with equality.

\section{Proof of Lemma \ref{thm:lemma_rank1}}\label{proof:lemma_rank1}

Since the set $\widehat{\Omega}$ is convex and compact, the boundary in direction $\mat{e}$, $\partial^{\mat{e}} \Omega, \mat{e} \in \mathcal{E},$  can be achieved using the Supporting Hyperplane theorem \cite[Theorem 1.5]{Tuy1998} by the following programming problem%
\begin{equation}\label{prb:origin}
\begin{split}
\text{maximize}& \quad \sum_{\ell = 1}^{K} \lambda_{\ell}e_\ell \bh_\ell^\H \bQ \bh_\ell\\
\text{subject to}& \quad \tr\bQ = 1, \quad \bQ \succeq 0,
\end{split}
\end{equation}

\noindent where $\mat{\lambda} \in \mat{\Lambda}$ defined in \eqref{eq:Lambda}. The objective in \eqref{prb:origin} can be written as%
\begin{equation}\label{eq:vonNeumann}
\begin{split}
\sum_{\ell = 1}^{K} \lambda_{\ell}e_\ell \bh_\ell^\H \bQ \bh_\ell &= \sum_{\ell = 1}^{K} \lambda_{\ell}e_\ell \tr{\bQ \bh_\ell \bh_\ell^\H}
= \text{tr}\Biggl({\bQ \underbrace{\sum_{\ell = 1}^{K} \lambda_{\ell}e_\ell \bh_\ell \bh_\ell^\H}_{\mat{Z}}}\Biggr) = \tr{\bQ \mat{Z}}
 \leq \sum_{\ell = 1}^{N} \mu_\ell(\bQ) \mu_\ell(\mat{Z})
\end{split}
\end{equation}
\noindent where the last inequality holds according to the von Neumann trace inequality of product of matrices \cite{Mirsky1959}. Define $\bQ = \bU_\bQ \Sigma_\bQ \bU_\bQ^\H$ and $\mat{Z} = \bU_{\mat{Z}} \Sigma_{\mat{Z}} \bU_{\mat{Z}}^\H$ where $\bU_\bQ$ and $\bU_{\mat{Z}}$ are unitary matrices. Then, the upper bound in \eqref{eq:vonNeumann} can be achieved by choosing $\bU_\bQ = \bU_{\mat{Z}}$. Hence, the problem in \eqref{prb:origin} can be equivalently written as
\begin{equation}
\begin{split}
\text{maximize}& \quad \sum_{\ell = 1}^{N} \mu_\ell(\bQ) \mu_\ell(\mat{Z})\\
\text{subject to}& \quad \sum_{\ell=1}^{N} \mu_{\ell}(\bQ) = 1, \quad \mu_\ell(\bQ) \geq 0, \quad \text{for all } \ell = 1,...,N.
\end{split}
\end{equation}

\noindent The solution of this problem is $\mu_{N}(\bQ) = 1$ and $\mu_{\ell}(\bQ) = 0$ if $\ell \neq N$. Thus, $\bQ$ is rank-1 and the transmit covariance matrices from $\widetilde{\mathcal{S}}$ achieve the boundary of the region $\widehat{\Omega}$.


\section{Proof of Theorem \ref{theo:1}}\label{proof:thm1}

As in the proof of Lemma \ref{thm:lemma_rank1}, the convex boundary of the set $\widehat{\Omega}$ in direction $\mat{e}$ can be characterized by the solution of the following programming problem

\begin{equation}\label{prb:opt_w}
\begin{split}
\text{maximize}& \quad \sum_{\ell=1}^K \lambda_\ell e_{\ell} \sabs{\mat{w}^H \mat{h}_{\ell}}\\
\text{subject to}& \quad \snorm{\bw} = 1,
\end{split}
\end{equation}
The objective function in \eqref{prb:opt_w} can be rewritten as
\begin{eqnarray}
y(\mat{w}) = \sum_{\ell=1}^K \lambda_\ell e_{\ell} \sabs{\mat{w}^H \mat{h}_{\ell}} = \mat{w}^H \Biggl(\underbrace{\sum_{\ell=1}^K \lambda_\ell e_{\ell} \mat{h}_{\ell} \mat{h}_{\ell}^H}_{\mat{Z}} \Biggr) \mat{w} \label{eq:mma}.
\end{eqnarray}
Note that the matrix $\mat{Z}$ in (\ref{eq:mma}) is not necessarily positive semidefinite because the directional vector $\mat{e}$ can contain negative components. However, it is Hermitian and therefore, the solution to \eqref{prb:opt_w} is the eigenvector which corresponds to the largest eigenvalue of $\mat{Z}$.

\section{Proof of Theorem \ref{thm:thm2} } \label{proof:thm2}

The proof is by contradiction. Assume that a power gain tuple $\mat{x}_k(\mat{w}_k \mat{w}_k^\H)$ is not on the boundary of the power gain region $\Omega_k$ and achieves a Pareto optimal point in the utility region. Since $\mat{x}_k(\mat{w}_k \mat{w}_k^\H)$ is not on the boundary of $\Omega_k$, we can find another power gain tuple on the boundary of the power gain region which dominates $\mat{x}_k(\mat{w}_k \mat{w}_k^\H)$ in the direction specified in \eqref{cases:e}. Accordingly, it is possible to increase the utility of at least one receiver without changing the utilities of the other receivers which follows from Assumption \ref{def:utility}. Thus, a contradiction is made on the Pareto optimality of $\mat{x}_k(\mat{w}_k \mat{w}_k^\H)$. Formally, assume
\begin{equation}\label{eq:assumption1}
\mat{x}_k(\mat{w}_k \mat{w}_k^\H) \notin \partial^{\mat{e}_k} \Omega_k,
\end{equation}%
\noindent and achieves a point on the Pareto boundary of $\mathcal{U}$, i.e.%
\begin{equation}\label{eq:assumption2}
(u_1(x_{1,1}(\mat{w}_1\mat{w}_1^\H),...,x_{T,1}(\mat{w}_T\mat{w}_T^\H)),...,u_K(x_{1,K}(\mat{w}_1\mat{w}_1^\H),...,x_{T,K}(\mat{w}_T\mat{w}_T^\H))) \in \mathcal{PB},
\end{equation}%
\noindent according to Definition \ref{def:1}. By assuming \eqref{eq:assumption1}, we can find another feasible beamforming vector $\mat{v}_k$ such that $\mat{v}_k \mat{v}_k^\H \in \mathcal{S}_k$ and achieves power gain tuple on the boundary of the power-gain region in direction $\mat{e}_k$ such that
\begin{equation}\label{eq:inequality}
x_{k,\ell}(\mat{w}_k \mat{w}_k^\H)e_{k,\ell} \leq x_{k,\ell}(\mat{v}_k\mat{v}_k^\H)e_{k,\ell}, \quad \text{for all } \ell \in \mathcal{K}.
\end{equation}
\noindent Next we distinguish two cases for the inequality in \eqref{eq:inequality} corresponding to $\ell \in \overline{\mathcal{K}}(k)$ or $\ell \in \underline{\mathcal{K}}(k)$.

\begin{enumerate}
\item Assume the inequality in \eqref{eq:inequality} is strict for $\ell \in \overline{\mathcal{K}}(k)$ with $e_{k,\ell} = +1$ as given in \eqref{cases:e}, then $x_{k,\ell}(\mat{w}_k\mat{w}_k^\H) < x_{k,\ell}(\mat{v}_k\mat{v}_k^\H)$. The power gains to all other receivers are to stay unchanged such that $x_{k,j}(\mat{w}_k\mat{w}_k^\H) = x_{k,j}(\mat{v}_k\mat{v}_k^\H)$ for $j \neq \ell$. Then,
    \begin{equation}
    u_\ell(x_{1,\ell}(\mat{w}_1\mat{w}_1^\H),...,x_{T,\ell}(\mat{w}_T\mat{w}_T^\H)) < u_\ell(x_{1,\ell}(\mat{w}_1\mat{w}_1^\H),...,x_{k,\ell}(\mat{v}_k\mat{v}_k^\H),...,x_{T,\ell}(\mat{w}_T\mat{w}_T^\H))
    \end{equation}
    \noindent holds according to property A in Assumption \ref{def:utility} in section \ref{sec:utility}. This result contradicts \eqref{eq:assumption2}.

\item Assume the inequality in \eqref{eq:inequality} is strict for $j \in \underline{\mathcal{K}}(k)$ with $e_{k,j} = -1$. Then, \eqref{eq:inequality} changes to $x_{k,j}(\mat{w}_k) > x_{k,j}(\mat{v}_k)$. Assuming $x_{k,\ell}(\mat{w}_k) = x_{k,\ell}(\mat{v}_k)$ for $\ell \neq j$, then
    \begin{equation}
    u_j(x_{1,j}(\mat{w}_1\mat{w}_1^\H),...,x_{T,j}(\mat{w}_T\mat{w}_T^\H)) < u_j(x_{1,j}(\mat{w}_1\mat{w}_1^\H),...,x_{k,j}(\mat{v}_k\mat{v}_k^\H),...,x_{T,j}(\mat{w}_T\mat{w}_T^\H))
    \end{equation}
    \noindent holds according to property B in Assumption \ref{def:utility} in section \ref{sec:utility}. This result contradicts \eqref{eq:assumption2}.

\end{enumerate}

\noindent Therefore, all points on the Pareto boundary of the utility region are achieved if each transmitter chooses its beamforming vectors to achieve the boundary of the gain-region in the direction specified in \eqref{cases:e}. These beamforming vectors are characterized in Theorem \ref{theo:1} in Section \ref{sec:NgeqK} and for the optimal power control in \eqref{cases:NlessK} in Section \ref{sec:NlessK}, which lead to the formulation in Theorem \ref{thm:thm2}.

\section{Proof of Lemma \ref{thm:zf}} \label{proof:zf}
The zero-forcing condition on the beamforming vector of transmitter $k$, denoted as $\bw_k^{ZF}, k \in \mathcal{K}$, is
\begin{eqnarray}\label{cases:zfcond}
	x_{k,\ell}(\bw_k^{ZF}) : \begin{cases} \geq 0 & k = \ell \\
	=0 & \text{otherwise} \end{cases}.
\end{eqnarray}
\noindent It is possible to fulfill the condition in \eqref{cases:zfcond} if $N_k \geq K$ and all channel vectors from transmitter $k$ to all receivers are linearly independent. Here, we give a direct proof. Assume the conditions $\lambda_{k,\ell} = 0$ for $k = \ell$ and $\lambda_{k,\ell} > 0$ for $k \neq \ell$ hold, we have to show that these conditions lead to beamforming vectors that satisfy \eqref{cases:zfcond}. In order to do this, we have to study the matrix in \eqref{eq:BF_pareto} whose eigenvector, corresponding to the largest eigenvalue, determines the used beamforming vector. Define the matrix $\mat{M}_k$, as
\begin{equation}\label{eq:MISO_Mk}
\begin{split}
\mat{M}_k &= \sum_{\ell=1}^K \lambda_{k,\ell} e_{k,\ell} \mat{h}_{k,\ell} \mat{h}_{k,\ell}^H = \underbrace{\lambda_{k,k} \bh_{kk} \bh_{kk}^\H}_{\mat{A}_k} - \underbrace{\sum\limits_{\ell \neq k} \lambda_{k,\ell}\bh_{k\ell} \bh_{k\ell}^\H}_{\mat{B}_k},
\end{split}
\end{equation}
\noindent where the direction vector $\mat{e}$ is specified as for the MISO IC application. According to the conditions $\lambda_{k,\ell} = 0$ for $k = \ell$ and $\lambda_{k,\ell} > 0$ for $k \neq \ell$, $\mat{A}_k$ in \eqref{eq:MISO_Mk} is equal to zero, and hence $\mat{M}_k$ is negative semidefinite. The largest eigenvalue of $\mat{M}_k$ is therefore zero since $N_k \geq K$, and we can write $({\sum\limits_{\ell \neq k} \lambda_{k,\ell}\bh_{k\ell} \bh_{k\ell}^\H}) \bw_k = \mat{0}.$ Since all channel vectors are linearly independent, the eigenvector associated with the largest eigenvalue produces zero gain on any of the interference channel vectors. It is clearly seen that if $\lambda_\ell=0$ for any $\ell \neq k$, then the largest eigenvector does not necessarily produce zero gain on this receiver.


\section{Proof of Corollary \ref{thm:cor1}} \label{proof:null_NE}

We prove that the gains achieved by the beamforming vectors in \eqref{eq:BF_null} are equal to the gains achived by the beamforming vectors given in \eqref{eq:BF_pareto}. Define the matrix $\mat{M}_k$, with the direction vector $e_{k,\ell}$ given in \eqref{cases:e}, as
\begin{equation}
\begin{split}
\mat{M}_k &= \sum_{\ell=1}^K \lambda_{k,\ell} e_{k,\ell} \mat{h}_{k,\ell} \mat{h}_{k,\ell}^H = \underbrace{\sum\limits_{\ell \in \mathcal{\overline{K}}(k)}\lambda_{k,\ell} \bh_{k\ell} \bh_{k\ell}^\H}_{\mat{A}_k} + \underbrace{\sum\limits_{\ell \in \mathcal{\underline{K}}(k)} -\lambda_{k,\ell}\bh_{k\ell} \bh_{k\ell}^\H}_{\mat{B}_k}
\end{split}
\end{equation}
\noindent The matrices $\mat{M}_k$, $\mat{A}_k$ and $\mat{B}_k$ are Hermitian matrices of size $N_k \times N_k$. The eigenvalues of $\mat{M}_k$ are real and we always consider them ordered in nondecreasing order, i.e., $\mu_1(\mat{M}_k) \leq \mu_2(\mat{M}_k) \leq ... \leq \mu_{N_k}(\mat{M}_k)$. $\mat{A}_k$ consists of sum of positive semidefinite matrices. Hence, $\mat{A}_k \succeq 0$ and $\rank{\mat{A}_k} \leq \abs{\mathcal{\overline{K}}(k)}$, i.e.,
\begin{equation}
0 \leq \mu_{\abs{\mathcal{\overline{K}}(k)} +1}(\mat{A}_k) \leq ... \leq \mu_{N_k}(\mat{A}_k), \text{ and } \mu_{1}(\mat{A}_k) = ... = \mu_{N_k - \abs{\mathcal{\overline{K}}(k)}}(\mat{A}_k) = 0.
\end{equation}%
%
%
$\mat{B}_k$ consists of the sum of the negative of positive semidefinite matrices. Hence, $\mat{B}_k \preceq 0$ and $\rank{\mat{B}_k} \leq \abs{\mathcal{\underline{K}}(k)}$, which leads to the following properties on the eigenvalues:
\begin{equation}
\mu_{1}(\mat{B}_k) \leq ... \leq \mu_{\abs{\mathcal{\underline{K}}(k)}}(\mat{B}_k) \leq 0, \text{ and } \mu_{\abs{\mathcal{\underline{K}}(k)}+1}(\mat{B}_k) =... = \mu_{N_k}(\mat{B}_k) = 0.
\end{equation}%
%
Next, we study the eigenvalues of $\mat{M}_k = \mat{A}_k + \mat{B}_k$. According to Weyl's inequality of the eigenvalues of the sum of Hermitian matrices \cite[Theorem 4.3.7]{Horn1985} the following properties are gained:
\begin{eqnarray}
\mu_{{N_k} - \abs{\mathcal{\overline{K}}(k)}}(\mat{M}_k) \leq \mu_{{N_k} - \abs{\mathcal{\overline{K}}(k)}}(\mat{A}_k) + \mu_{{N_k}}(\mat{B}_k) = 0 \\
\mu_{\abs{\mathcal{\underline{K}}(k)}+1}(\mat{M}_k) \geq \mu_{1}(\mat{A}_k) + \mu_{\abs{\mathcal{\underline{K}}(k)}+1}(\mat{B}_k) = 0
\end{eqnarray}%

\noindent The eigenvalues of $\mat{M}_k$ are ordered in nondecreasing order. Therefore, the following eigenvalues of $\mat{M}_k$ are always equal to zero: $\mu_{\abs{\mathcal{\underline{K}}(k)}+1}(\mat{M}_k) = ...= \mu_{N_k - \abs{\mathcal{\overline{K}}(k)}}(\mat{M}_k) = 0.$ In addition, the smallest $\abs{\mathcal{\underline{K}}(k)}$ eigenvalues of $\mat{M}_k$ are nonpositive.

If the dimension of space is larger than the number of receivers, i.e., $N_k \geq \abs{\mathcal{K}}$, then there would be at least $N_k - \abs{\mathcal{\underline{K}}(k)} - \abs{\mathcal{\overline{K}}(k)}$ eigenvalues of $\mat{M}_k$ that are zero. For the eigenvectors corresponding to those eigenvalues, the eigenvalue equation is written as $({\sum\limits_{\ell \in \mathcal{\overline{K}}(k)}\lambda_{k,\ell} \bh_{k\ell} \bh_{k\ell}^\H + \sum\limits_{\ell \in \mathcal{\underline{K}}(k)} -\lambda_{k,\ell}\bh_{k\ell} \bh_{k\ell}^\H})\bv_{i} = \mat{0},$ for all $i = \abs{\mathcal{\underline{K}}(k)}+1,...,N_k - \abs{\mathcal{\overline{K}}(k)}$. Then, for all $\ell \in \mathcal{K}$,
\begin{equation}\label{eq:eigvect_zero}
\pp{\lambda_{k,\ell} \mat{h}_{k\ell}\mat{h}_{k\ell}^\H}\bv_{i} = 0, \quad \text{for all } i = \abs{\mathcal{\underline{K}}(k)}+1,...,N_k - \abs{\mathcal{\overline{K}}(k)}.
\end{equation}


The set of eigenvectors of $\mat{M}_k$, $\{\bv_1,...,\bv_{N_k}\}$, form an orthonormal set, i.e. $\norm{\bv_i} = 1$ for all $i = 1,...,N_k$ and $\bv_i^\H\bv_j = 0$ for $i \neq j$. Therefore, we can write $\sum_{\ell = 1}^{N_k} \bv_\ell\bv_\ell^\H = \bI$, which gives
\begin{equation}\label{eq:sum_cov_eig}
\begin{split}
\bv_{N_k}(\mat{\lambda}) \bv_{N_k}^\H(\mat{\lambda}) & = \bI - \sum_{\ell = 1}^{N_k-1} \bv_\ell(\mat{\lambda})\bv_\ell^\H(\mat{\lambda}) = \bI - G_k(\lambda)G_k^\H(\lambda) = \mat{\Pi}_{G_k(\lambda)}^\perp,
\end{split}
\end{equation}%
\noindent where $G_k(\lambda) = \left[\mat{v}_1(\mat{\lambda}),...,\mat{v}_{N_k-1}(\mat{\lambda})\right].$ Let the matrix $Z_k(\mat{\lambda})$ consist of the eigenvectors of $G_k(\lambda)$ excluding the eigenvectors that satisfy \eqref{eq:eigvect_zero}, i.e.,
\begin{equation}
Z_k(\mat{\lambda}) = \left[\mat{v}_1(\mat{\lambda}),..., \mat{v}_{\abs{\mathcal{\underline{K}}(k)}}(\mat{\lambda}), \mat{v}_{N_k - \abs{\mathcal{\overline{K}}(k)} +1}(\mat{\lambda}),...,\mat{v}_{N_k-1}(\mat{\lambda})\right],
\end{equation}
\noindent then for any $\mat{g} \in \mathbb{C}^{N_k}$ we can write%
\begin{align}
\sabs{\mat{g}^\H\frac{\mat{\Pi}_{Z_k(\lambda)}^\perp \mat{h}_{k \ell}}{\norm{\mat{\Pi}_{Z_k(\lambda)}^\perp \mat{h}_{k\ell}}}} &= \sabs{\mat{g}^\H\frac{ \mat{\Pi}_{G_k(\lambda)}^\perp \mat{h}_{k\ell}}{\norm{\mat{\Pi}_{G_k(\lambda)}^\perp \mat{h}_{k\ell}}}} = \sabs{\mat{g}^\H\frac{\bv_{N_k}(\mat{\lambda})\bv_{N_k}^\H(\mat{\lambda})\mat{h}_{k\ell}}{\norm{\bv_{N_k}(\mat{\lambda})\bv_{N_k}^\H(\mat{\lambda}) \mat{h}_{k\ell}}}}= \sabs{\mat{g}^\H\bv_{N_k}(\mat{\lambda})},
\end{align}
\noindent where $\ell \in \mathcal{K}$. Hence, the same power gains are achieved with $\frac{\mat{\Pi}_{Z_k(\lambda)}^\perp \mat{h}_{k \ell}}{\norm{\mat{\Pi}_{Z_k(\lambda)}^\perp \mat{h}_{kk}}}$ as with $\bv_{N_k}(\mat{\lambda})$.

\bibliographystyle{IEEEtran}
\bibliography{references}

\end{document}